\documentclass[a4paper]{aa}

\usepackage{graphicx}
\usepackage{array}
\usepackage{txfonts}
\usepackage{caption}
\usepackage{lipsum}
\usepackage{hyperref}
\hypersetup{colorlinks, allcolors=blue}
\usepackage{url}
\usepackage{natbib}
\usepackage{breqn}
\usepackage{diagbox}
\usepackage{cuted} 
\usepackage{lscape}
\usepackage[mathscr]{euscript}
\usepackage{placeins}
\usepackage{titlesec}
\usepackage{comment}

\begin{document} 

   \title{The Solar System through dust maps}
   \subtitle{Encounter statistics with interstellar gas}
\author{  E. Maconi\inst{1},
          J. Alves\inst{1},
          S. Hutschenreuter\inst{1},
          S. Ratzenb\"ock \inst{1,2,3},
          M. Pantaleoni\inst{1},
          C. Zucker\inst{3},
          C. Levy\inst{3,4},
          \and
          G. Edenhofer\inst{5}
          }
\institute{
        University of Vienna, Department of Astrophysics, T\"urkenschanzstraße 17, 1180 Wien, Austria. \\ \email{efrem.maconi@univie.ac.at}
        \and
        University of Vienna, Research Network Data Science at Uni Vienna, Kolingasse 14-16, 1090 Vienna, Austria
        \and
        Center for Astrophysics -- Harvard \& Smithsonian, 60 Garden Street, Cambridge, MA 02138, USA
        \and
        Department of Physics, Harvard University, 17 Oxford St, Cambridge, MA 02138, USA
        \and 
        Independent Researcher 
        }

\date{Received ...; accepted ...}

\titlerunning{The Solar System through dust maps}
\authorrunning{E.~Maconi et al.}

\abstract
    {During its orbit around the Milky Way, the Solar System crosses different Galactic environments, including dense regions of the interstellar medium (ISM). These encounters increase the interstellar dust influx, leave detectable signatures in terrestrial geological records, and affect the structure of the heliosphere.}
    {This study aims to characterize the Galactic environments crossed by the Solar System along its orbit. For the local ISM, we estimate the frequency and duration of crossings, the fraction of time spent above different gas density thresholds (residence fraction), and how the coupling between the Solar vertical motion and the vertical stratification of the ISM influences these statistics. We also quantify the associated gas influx and dust exposure.}
    {We use \textit{Gaia}-based 3D dust maps of the Solar neighborhood, which provide some of the most detailed reconstructions currently available of the local ISM structure. We sample these maps by integrating the Solar System along its Galactic orbit while rigidly translating the maps. From the resulting density time series, we derive the encounter statistics and estimate the corresponding gas and dust fluxes.}
    {We find that the Solar System spends most of its time in low density environments, while crossings of dense gas become rapidly rarer with increasing density threshold. For $n_{\rm H}\geq1\,{\rm cm}^{-3}$, the crossing frequency is above $600\,{\rm Gyr}^{-1}$, whereas thresholds around $200\,{\rm cm}^{-3}$ are crossed once per Gyr. 
    The duration of individual contiguous crossings is typically of the order of several tens of kyr and decreases toward higher density thresholds. However, the inferred durations soon become nearly constant, reflecting the finite spatial resolution of the maps rather than a physical saturation of cloud sizes. 
    The Sun spends $\sim18$ to $47\%$ of the sampled time above $1\,{\rm cm}^{-3}$, and $\sim1$ to $10\%$ above $10\,{\rm cm}^{-3}$. We also find that the residence fraction can be reproduced by combining the vertical volume filling factor of gas with the fraction of time spent by the Sun at each height. Within the current vertical range sampled by the Sun, we find no clear enhancement of the crossing statistics during midplane passages, suggesting that Galactic midplane crossings do not necessarily lead to periodic perturbations of the Solar System. Finally, about $50\%$ of the gas influx is contributed by gas with $n_{\rm H}\lesssim3-4\,{\rm cm}^{-3}$, and about $90\%$ by gas with $n_{\rm H}\lesssim48\,{\rm cm}^{-3}$.}
    {Our results provide a first order, data driven estimate of the encounter statistics with the ISM, gas influx, and dust exposure experienced by the Solar System along its Galactic orbit.} 

   \maketitle
%

\section{Introduction}\label{sect:intro}

During its orbit around the center of the Milky Way, the Solar System crosses the interstellar medium (ISM), encountering  environments characterized by varying gas and dust densities, elemental composition, temperatures, ionization states, and velocities \citep[see e.g.,][]{Myers1978,Frisch2011,Klessen2016}.
The ISM traversed by the Sun has received significant scientific attention for several reasons. For example, the density of the surrounding ISM and its relative velocity with respect to the Sun affect the size and structure of the heliosphere \citep[see e.g.,][]{Zank1999,Muller2006,Miller2022}, which in turn plays a key role in modulating the Galactic cosmic-ray flux \citep[see e.g.,][]{Webber2013,Stone2013}. 
The encountered ISM also gives rise to an influx of interstellar dust into the Solar System, as first detected by the \textit{Ulysses} spacecraft \citep[see e.g.,][]{Grun1993,Gruen1994,Kruger2015} and later confirmed by \textit{Galileo} and \textit{Cassini} \citep[see e.g.,][]{Baguhl1996,Altobelli2003}. Evidence that such material can enter the Earth's atmosphere was presented by \citet{Taylor1996}. Encounters with different ISM environments along the Sun's Galactic orbit have also been linked to anomalies in geological records \citep[see e.g.,][]{Knie1999,Koll2019,Wallner2020}, possible gravitational influences on the Oort cloud \citep[see e.g.,][]{Hut1985,Jakubik2009,Torres2019}, and, more tentatively, Earth's climate \citep[see e.g.,][]{Shapley1921,Hoyle1939,Pavlov2005,Opher2024}.

At present, the Solar System resides within the Local Bubble, a large ($R\sim100-200\,\mathrm{pc}$), hot ($T\sim10^6\,\mathrm{K}$), and low-density ($n\sim0.01\,\mathrm{cm^{-3}}$) cavity of the ISM \citep[see e.g.,][]{coxANDreynolds1987,Galeazzi2014,Breitschwerdt2016,Zucker2022,ONeill2024}.
More specifically, the Sun is currently crossing a complex of diffuse ($n_{\mathrm{H}} \sim 0.1-0.2 \, {\mathrm{cm^{-3}}}$), warm ($T\sim5,000 - 10,000 \, \mathrm{K}$), and partially ionized clouds that are located inside the Local Bubble. These clouds, known as the Cluster of Local Interstellar Clouds \citep[CLIC; e.g.,][]{Lallement1992,Redfield2008,Vannier2025}, are thought to have formed from a supernova occurring within the pre-evacuated Local Bubble \citep{Zucker2025}, are associated to an outflow originating from the Scorpius-Centaurus OB association \citep{Piecka2024}, and 
have also been linked to the recent deposition of $^{60}$Fe detected in Antarctic snow and ice-cores \citep{Koll2019,Koll2026}. 

In its recent ($\lesssim 20$ Myr) Galactic orbit, the Sun has not always resided within the Local Bubble. Approximately 14 Myr ago, while still outside this cavity, whose expansion started around the same epoch \citep[see e.g.,][]{Zucker2022}, the Sun was traversing the Radcliffe Wave, a coherent $\sim$2.7 kpc-long massive gaseous structure \citep{Alves2020}, in the Orion region \citep{Maconi2025}. The Solar System may subsequently have crossed the Orion–Eridanus superbubble \citep[see e.g.,][]{Schulreich2023} before entering the Local Bubble about 5-6 Myr ago \citep[see e.g.,][]{Zucker2022}. Since then, it has traveled through the Local Bubble interior, possibly crossing the Lynx cloud \citep[][]{Opher2024}, and entered the CLIC between $\sim240,000-160,000$ yr ago \citep[][]{Vannier2025}, reaching its current location.

\begin{figure*}[!htbp]
    \centering
    \includegraphics[width=0.8\textwidth]{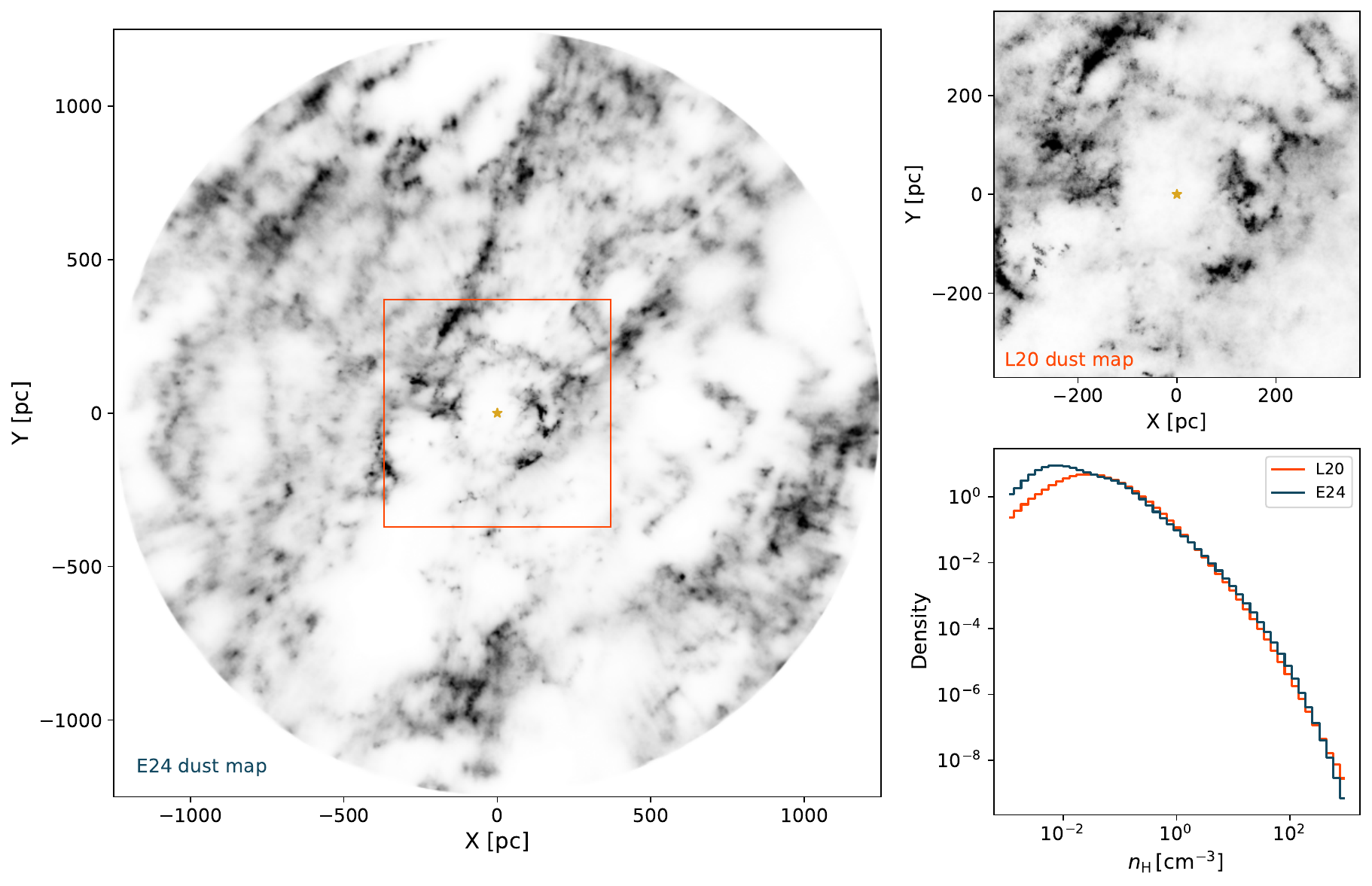}
    \caption{Overview of the dust maps used in this work. \textit{Left panel}: E24 dust map projected onto the \textit{xy}-plane. The red-orange square highlights the extent of the L20 map. \textit{Top-right panel}: L20 dust map projected on the \textit{xy}-plane. The yellow star marks the current position of the Solar System. \textit{Bottom-right panel}: probability density functions of the gas density for both maps.}
    \label{fig:maps_overview}
\end{figure*}

Reconstructing the Solar trajectory further back in time becomes increasingly uncertain due to potential gravitational scattering and the difficulty to reconstruct past gas structures. An alternative approach is to statistically characterize the environments encountered by the Sun, for example by estimating the frequency of crossings at a given density. 
Different studies have addressed this question using either analytical estimates applied to observational constraints \citep[see e.g.,][]{Talbot1977} or numerical simulations \citep[see e.g.,][]{Kokaia2019}.

In this paper, we aim to characterize the ISM crossed by the Sun using the three-dimensional (3D) dust maps of the Solar neighborhood by \citet{Leike2020} and \citet{Edenhofer2024}. These maps, together with other reconstructions available in the literature, were made possible by the European Space Agency's \textit{Gaia} mission \citep{Gaia2016} together with the advent of deep wide-field photometric surveys \citep[e.g., Pan-STARRS1,][]{Chambers2016}, and the advancement of statistical techniques \citep[see e.g.,][]{Edenhofer2024_nifty}. Collectively, these reconstructions  represent the first detailed 3D view of the Solar neighborhood, revealing previously unknown Galactic structures \citep[see e.g.,][]{Lallement2019,Alves2020,Kormann2026}. 

Our approach consists of rigidly translating the maps along a circular orbit and sampling their density fields along the trajectory of a point particle moving on simulated Solar orbits. From the resulting time series, we estimate the frequency and duration of passages through different ISM density regimes, the fraction of time spent above different density thresholds, and the associated gas and dust exposure.
The main advantage of this approach is that the sampled density field is based on observationally informed 3D dust reconstructions of the local ISM. Its main limitation is that the maps provide a present day snapshot of a constantly evolving ISM, cover a local portion of the Galaxy, and are not very sensitive to gas densities above $n_{\rm{H}} \sim100-200 \, \rm{cm}^{-3}$.

This paper is organized as follows. In Sect.~\ref{Sect.:Data}, we describe the dust maps used for this study; in Sect.~\ref{Sect.:Methods}, we present the integration method and maps sampling procedure, define the encounter statistics and influx estimates; and in Sect.~\ref{Sect:results}, we present and discuss our results. Finally, we summarize our results in Sect.~\ref{Sect.:conclusions}.


\section{Data}\label{Sect.:Data}

\subsection{Dust maps}

We use the 3D interstellar dust extinction maps of the Solar neighborhood by \citet[][hereafter L20]{Leike2020} and \citet[][hereafter E24]{Edenhofer2024} to characterize the ISM traversed by the Solar System along its Galactic orbit. The two maps are not combined into a single density field, but are analyzed independently throughout this work. We chose these maps because they currently offer the best distance resolution \citep[see also the discussion in][]{Zucker2025_dustMap}, which is essential for resolving high-density regions. Additionally, we used the two maps as they differ in spatial extent, resolution, and spatial discretizations, allowing us to test how these quantities affect the results.

\begin{figure*}
    \centering
    \includegraphics[width=0.85\textwidth]{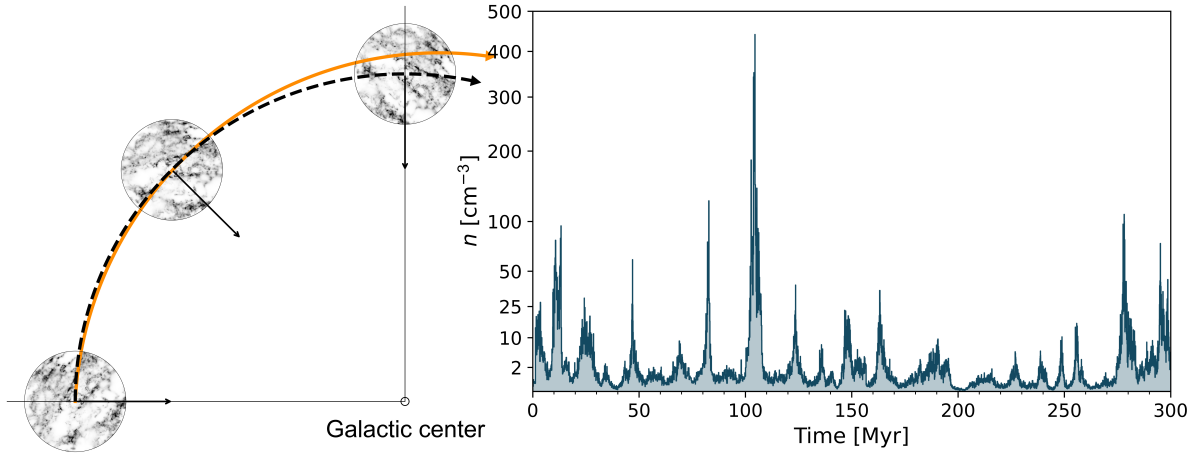}
    \caption{
    Overview of the setup used to sample the maps and example output from this procedure.
    \textit{Left panel}: Schematic representation of the rigid translation of the E24 dust map, overlaid with the solar orbit (solid dark orange line). The map follows a circular Galactic orbit (dashed black line) and remains oriented toward the Galactic center. The same setup is used for the L20 map.
    \textit{Right panel}: Example of the density encountered by the Solar System along its trajectory using this setup. The y-axis is shown on a square-root scale. The example corresponds to one realization using the E24 map; similar behavior is obtained with the L20 map.}
    \label{fig:setup_overview}
\end{figure*}

\subsubsection{Leike et al. (2020)}

The L20 dust map is based on the stellar parameter catalog by \citet{Anders2019}, which combines observational data from \textit{Gaia} DR2, AllWISE, Pan-STARRS, and 2MASS. 
The map is based on a Cartesian grid centered on the Sun, with a voxel resolution of 1 pc and a reported functional resolution of 2 pc. The grid extends 740 pc in the \textit{xy}-plane and 540 pc in the \textit{z}-direction. 

The L20 map is provided as a set of 12 posterior samples of the extinction density. We used all the 12 samples for our analysis. 
The map is expressed in terms of differential extinction, given as the \textit{Gaia} G-band optical depth, $\tau_G$, per parsec. To convert the differential extinction to gas volume density, we follow the approach of \citet{Zucker2021} and provide further details in Appendix~\ref{Sect:Ext_to_Dens}.

\subsubsection{Edenhofer et al. (2024)}

The E24 dust map is based on the catalog of stellar distance and extinction estimates by \citet[][hereafter ZGR23]{Zhang2023}, which is primarily based on \textit{Gaia} BP/RP spectra from \textit{Gaia} DR3 \citep{Gaia2023}. 
E24 modeled the 3D dust extinction within 1.25 kpc of the Sun in the HEALPix \citep{Gorski2005} spherical coordinate system, with 516 logarithmically spaced distance bins from 69 pc to 1250 pc, using the {\tt NIFTy} statistical library \citep{Edenhofer2022,Edenhofer2024_nifty}. The map has an angular resolution of $13.7'$ ($N_{\mathrm{side}}=256$) and a distance resolution ranging from 0.4 pc at 69 pc to 7 pc at 1.25 kpc. Within 180 pc from the Sun the distance resolution is 1 pc or better, and remains below 2 pc out to 360 pc. However, these values describe only the binning of the map, rather than its effective spatial resolution. The effective resolution remains on parsec scales, but depends on distance and on the number of stars constraining the reconstruction in each region.
The E24 map excludes the innermost 69 pc around the Sun, where extremely low extinction and previous reconstructions \citep[][]{Leike2020} indicate a negligible dust content.

E24 reconstructed 12 samples of the inferred 3D dust extinction distribution, along with the posterior mean. In our analysis, we used the 12 samples.
The reconstruction is provided in terms of the dimensionless extinction defined in ZGR23. As before, we refer to Appendix~\ref{Sect:Ext_to_Dens} for the conversion of the map to gas volume density values. 

The authors also provide the E24 map interpolated onto a Cartesian grid, together with the interpolation script. However, we perform our analysis on the original HEALPix grid. This choice avoids additional smoothing of high-density regions introduced by the interpolation. Moreover, the Cartesian interpolation does not fully remove the effect of the original distance-dependent bin sizes, since the interpolated grid still retains an imprint of the underlying HEALPix discretization.

\hfill \break

For both maps, the conversion procedure described in Appendix~\ref{Sect:Ext_to_Dens} yields the total volume density of hydrogen nuclei, independent of ISM phase \((n = n_{\mathrm{H\,I}} + 2 \, n_{\mathrm{H_2}})\).
The L20 and E24 maps are each subject to different sources of uncertainty, which propagate into our crossing statistics. We refer to the respective papers for a detailed discussion. Additional limitations relevant to our analysis include the potential underestimation of extinction toward dense clouds, due to the lack of stellar data at high densities to constrain the 3D reconstruction \citep[see also][]{Zucker2021} and, for the E24 map, the increasing volume bin size with increasing distance. Additionally, converting dust extinction into gas volume density introduces an additional uncertainty, as the dust-to-gas conversion factor is not universal and may vary with the local ISM conditions \citep[see e.g.,][]{ONeill2026}. These limitations will impact our results, as later discussed.

In Fig.~\ref{fig:maps_overview}, the projections of the two maps onto the \textit{xy}-plane are shown for one representative posterior sample each.
The L20 map captures the local ISM within 370~pc of the Sun, and local structures such as the Local Bubble are resolved in details.
However, due to its limited extent, it cannot unveil the large-scale organization of the ISM, including kpc-long structures such as the Radcliffe Wave \citep{Alves2020} and other superclouds \citep{Kormann2026}, which instead become recognizable in the E24 map.
In the same figure, we also show the probability density functions of the gas density of the maps. They are broadly consistent over most of the range relevant for the encounter statistics analyzed here. Differences appear mainly at very low densities, $n_{\rm H}\lesssim10^{-2}\,{\rm cm}^{-3}$, where the reconstructions are more uncertain.

\section{Methods}\label{Sect.:Methods}

\subsection{Integration of the orbits}\label{sect:methods-orbitInt}

To derive the crossing statistics and gas flux, we integrated the orbit of the Solar System together with that of the map. At each time step, the map is rigidly translated along a circular orbit at the Sun's current Galactocentric radius, with its orientation fixed toward the Galactic center, in order to preserve the orientation of the Galactic structures. This is not intended to model the full dynamical evolution of the ISM, but rather to account for the mean rotation of the local Galactic disk. In Appendix~\ref{sect:ValidationTests}, we also tested the case of a random map orientation. 
The setup described above is illustrated schematically in Figure~\ref{fig:setup_overview}. Throughout this work, when referring to the Sun or the Solar System in the context of the orbit integration, we mean a test particle initialized with the given initial conditions and integrated in a Milky Way-like gravitational potential.

For the orbits integration, we used the Galactic dynamics package {\tt galpy} \citep{Bovy2015}. We adopted the {\tt MWPotential2014} model for the Milky Way potential which includes a bulge, disk, and dark-matter halo component \citep[see][for details]{Bovy2015}.
The velocity of the LSR was set to $v_{\mathrm{LSR}} = 220 \, \mathrm{km}\,\mathrm{s}^{-1}$ \citep[see e.g.,][]{Bovy2012}.
The map follows a circular orbit at $R = R_{\odot}$ in the Galactic midplane, with a velocity relative to the Galactic center of $(U, V, W) = (0, 220, 0)\,\mathrm{km\,s^{-1}}$.

In the reference integration, we positioned the Sun at the Galactocentric radius of $R_{\odot} = 8.122 \, \mathrm{kpc}$ \citep{GRAVITY2018} and at a height above the disk $z_{\odot}= 20.8 \, \mathrm{pc}$ \citep{Bennett2019}. The solar velocity relative to the LSR was set to ($U_\odot,\, V_\odot,\, W_\odot$) = (11.1, 12.24, 7.25)$\,\mathrm{km}\,\mathrm{s}^{-1}$ \citep{Schoenrich2010}.

To study how the ISM structure and the vertical motion of the Sun affect the crossing statistics, we also performed a set of controlled experiments. In these tests, we modified the initial $z$ and $v_z$ values of the Sun to sample different maximum heights, or modified the ISM density field within a given height range by randomly shuffling the density values among the map cells. This preserves the density distribution, but removes the spatial correlations. These experiments are described in Sect.~\ref{sect:results-ISM_stratification}. 
Additional tests using different initial conditions are presented in Appendix~\ref{sect:ValidationTests}.

In all cases, the integrations were performed such that the map is sampled by the Sun for at least $2\,\mathrm{Gyr}$.\footnote{This because the maps have finite spatial extent and the Sun can leave and later re-enter the map volume during the integration.} We adopted a timestep of $0.001\,\mathrm{Myr}$, ensuring that the corresponding spatial sampling is always smaller than the minimum maps resolution of $0.4\,\mathrm{pc}$. This results in a typical spatial step of $\sim0.2\,\mathrm{pc}$
in the Galactocentric reference frame and of $\sim0.05\,\mathrm{pc}$ relative to the maps.

To account for the statistical uncertainties in the Sun’s position and velocity, we repeated the orbit integration 100 times, each with different initial conditions drawn via Monte Carlo sampling of the associated uncertainties, reported in the referenced literature. Statistical uncertainties in the dust maps were addressed by applying this procedure to each of the 12 samples. In all cases, we considered the center of the maps to be positioned at the Galactocentric height of $20.8\,{\rm pc}$, assuming the Sun's height by \citet{Bennett2019} as reference.

\subsection{Sampling of the dust maps}\label{sect:methods-mapSampling}

We tracked the positions of the Sun and the map throughout the integration. At each time step for which the Sun is located inside the map boundaries, we recorded its position and velocity relative to the map together with the gas density. This procedure produces a time series of sampled densities, an example of which is shown in Fig.~\ref{fig:setup_overview}. 
We then considered density thresholds ($n_{\rm th}$) between $1$ and $500\,{\rm cm}^{-3}$ and derived three main encounter statistics, described in \S\ref{sect:methods-crossStats}, and gas flux and mass accretion estimates, as outlined in \S\ref{sect:methods-gasFlux}.

\subsubsection{Encounter statistics}\label{sect:methods-crossStats}

For each density threshold, we identified contiguous time intervals during which the encountered density satisfies $n \geq n_{\rm th}$. A crossing is defined as an event beginning when the encountered density rises above the threshold and ending when it drops below the threshold again.
We estimated the following encounter statistics: crossing frequency, $f_{\rm cross}(\geq n_{\rm th})$, crossing duration, $\tau_{\rm cross}(\geq n_{\rm th})$, and residence time fraction, $\eta_{\rm res}(\geq n_{\rm th})$. Similar encounter statistics have been used in previous studies \citep[see e.g.,][]{Talbot1977,Kokaia2019,Pfalzner2020}, although their precise definitions are not always stated explicitly. We define the quantities used in this work as follows.

The crossing frequency is computed to quantify how often the Solar System encounters regions above a given density. 
For each threshold, we counted the number of crossings, $N$, above a density threshold. The crossing frequency is then defined as
\begin{equation}
f_{\rm cross}(\geq n_{\rm th}) =
\frac{N(\geq n_{\rm th})}{T_{\rm inside}},
\end{equation}
where $N(\geq n_{\rm th})$ is the number of crossings above the threshold and $T_{\rm inside}$ is the total time for which the Solar orbit is inside the map volume.
For each dust map sample, we combined the 100 orbit realizations by summing both the number of crossings and the corresponding sampled time. This gives one long-time estimate of the crossing frequency for each dust map sample. We then reported the median over the 12 dust map samples, together with the 16th and 84th percentiles.

The crossing duration quantifies the typical time spent in a single contiguous encounter above a given density threshold.
We measured the duration of each individual crossing, $\Delta t_{{\rm cross},i}(\geq n_{\rm th})$, across the dust map samples and orbit realizations. The characteristic crossing duration, $\tau_{\rm cross}(\geq n_{\rm th})$, is defined as the median of this distribution, and the 16th and 84th percentiles are used to describe the spread of individual crossing durations.

The residence time fraction quantifies the fraction of time that the Sun spends in regions whose density exceeds a given value. It is defined as
\begin{equation}
\eta_{\rm res}(\geq n_{\rm th}) =
\frac{T_{\rm inside}(\geq n_{\rm th})}{T_{\rm inside}},
\end{equation}
where $T_{\rm inside}(\geq n_{\rm th})$ is the total time spent above the threshold. Following the same procedure as for the crossing frequency, we report the median and the 16th to 84th percentile range over the 12 dust map samples.

These quantities, $f_{\rm cross}$, $\tau_{\rm cross}$, and $\eta_{\rm res}$, are not independent. To first order, they are related by
\begin{equation}\label{eq:crossStatistics-relation}
 f_{\rm cross} \sim \frac{\eta_{\rm res}}{\tau_{\rm cross}}.
\end{equation}
They therefore provide complementary descriptions of the encounter history and are sensitive to different properties of the density field and of the orbit, as discussed in Sect.~\ref{Sect:results}.

We also computed the volume filling factor of the gas,
\begin{equation}
f_{\rm Vol}(\geq n_{\rm th}) = \frac{V( \geq n_{\rm th})}{V_{\rm tot}},
\end{equation}
where $V( \geq n_{\rm th})$ is the volume occupied by gas above the threshold and $V_{\rm tot}$ is the total spatial volume considered. When studying the vertical structure of the maps, we computed the same quantity in bins of $|z|$. The values of $f_{\rm Vol}$ shown later in the figures correspond to the median over the 12 dust map samples.

\subsubsection{Gas flux and interstellar dust accretion estimates}\label{sect:methods-gasFlux}

Our setup allows us to estimate, to first order, the time-averaged gas flux encountered by the heliosphere, $\Phi_{\rm gas}^{\rm tot} \, [{\rm cm^{-2} \cdot  yr^{-1}}]$, as
\begin{equation}\label{eq:mean_gas_flux}
    \Phi_{\rm gas}^{\rm tot} = \frac{1}{T_{\rm inside}} \int_0^{T_{\rm inside}} n(t) \, v_{\rm rel}(t) \, {\rm d}t,
\end{equation}
where $n(t)$ and $v_{\rm rel}(t)$ are the gas density and relative velocity between the Sun and the map at the integration time $t$, respectively. 
Since the dust maps are static, $v_{\rm rel}(t)$ only accounts for the relative motion between the Solar orbit and the translated map, and does not include intrinsic gas motions within the ISM.
We also computed the relative flux contribution of different gas density $\Phi_{\rm gas}(\geq n_{\rm th})$ to the total flux.

As an order of magnitude calculation, we convert the gas flux into a simple estimate of the interstellar dust mass accretion rate onto Earth, $\dot M_{\rm dust} \, [{\rm kg \cdot  yr^{-1}}]$. 
We approximate this rate as
\begin{equation}\label{eq:mass_rate_full_eq}
    \dot M_{\rm dust} = \frac{1}{T_{\rm inside}} \int_0^{T_{\rm inside}}  \, \rho_{\rm dust}(t) \, v_{\rm rel}(t) \, \sigma_{\rm eff}(t) \, \epsilon(t) \, {\rm d}t,
\end{equation}
where $\rho_{\rm dust}(t)$ is the dust mass density, $\sigma_{\rm eff}(t)$ the effective cross section of the Earth, and $\epsilon(t)$ a dimensionless efficiency factor. This factor accounts for processes that regulate the delivery of interstellar dust to Earth, including gravitational focusing, heliosphere filtering, and grain-size- and charge-dependent transport of interstellar dust. The modeling of these processes is beyond the scope of this work. For a more in depth discussion, we refer the reader to \citet{Slavin2012}, \citet{Kruger2015}, \citet{Godenko2024}, and references therein.

Equation~\ref{eq:mass_rate_full_eq} can be simplified by additionally assuming that the cross section and filtering factor do not depend on time, and that the ISM is well mixed. From this, the dust mass density can be derived as $\rho_{\rm dust}(t) = D \,\mu_{\rm H} \, m_{\rm H} \, n(t)$, with $D$ being the dust-to-gas mass ratio, $\mu_{\rm H}$ the mass contribution from helium and metals per hydrogen nucleus, and $m_{\rm H}$ the hydrogen mass. 
Equation~\ref{eq:mass_rate_full_eq} finally reduces to
\begin{equation}\label{eq:mass_rate_full_eq_simplyfied}
    \dot M_{\rm dust} = {D \,\mu_{\rm H} \, m_{\rm H} \, \sigma_{\rm eff} \, \epsilon} \,\Phi_{\rm gas}^{\rm tot},
\end{equation}
which is the first order approximation we used in this work. For the parameters, we assumed $D=0.01$, $\mu_{\rm H}=1.4$, $m_{\rm H}=1.67\cdot 10^{-27}\,\rm{kg}$, and $\sigma_{\rm eff}=\sigma_{\oplus}$, with $\sigma_{\oplus}$ being the geometrical cross section of the Earth.


\begin{figure}[!htbp]
    \centering
    \includegraphics[width=0.45\textwidth]{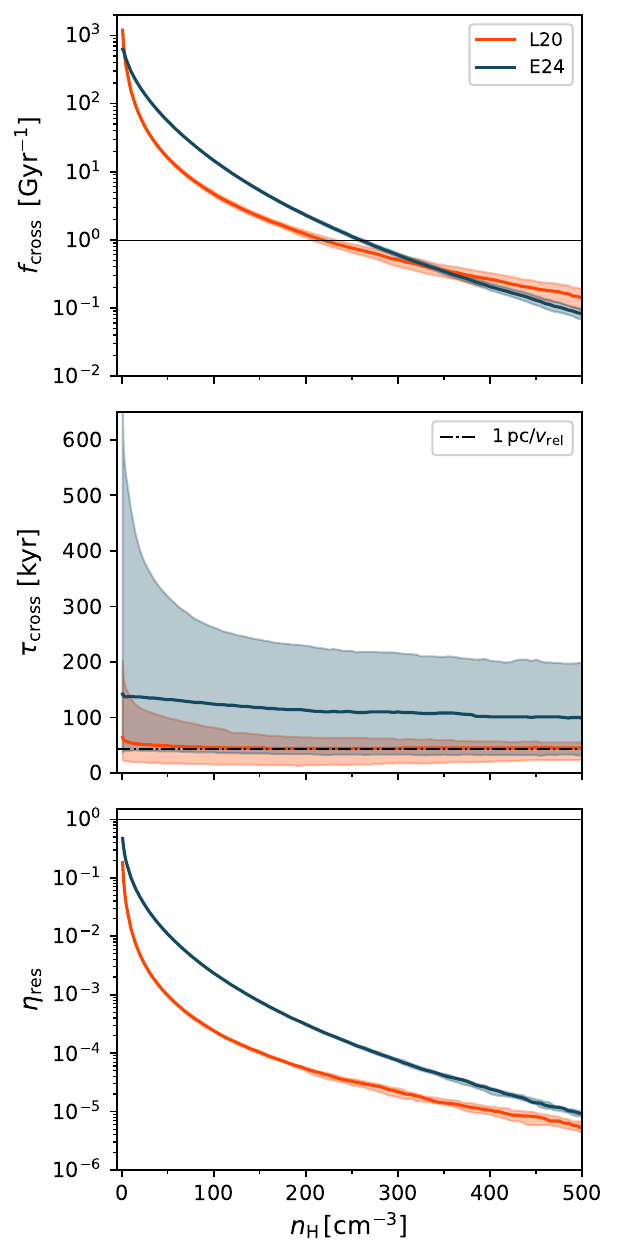}
    \caption{Encounter statistics as a function of gas density. The results regards a stellar particle initialized with the present-day phase-space coordinates of the Solar System.
    \textit{Top panel}: Crossing frequency per Gyr. The horizontal black line represents one crossing per Gyr.
    \textit{Mid panel}: Duration of individual crossings. The horizontal dashed line represent the time required to cross $1\,\rm{pc}$ distance with the average relative velocity between the Sun and the map.
    \textit{Bottom panel}: Residence time fraction. The horizontal black line represent a residence time of 100\%.
    The values reported represent the medians, as described in Sect. \ref{sect:methods-crossStats}. The shaded areas represent the 16th and 84th percentile range.
    }  
    \label{fig:crossing_statistics}
\end{figure}

\begin{table}
	\centering
    \renewcommand{\arraystretch}{1.5}
	\caption{Encounter statistics evaluated at representative density thresholds for both maps. The selected thresholds sample the full density range shown in Fig.~\ref{fig:crossing_statistics}, from diffuse gas to the rarest dense structures resolved by the maps. The table reports only central median values for readability.}
	\label{tab:crossing_statistics}
	\begin{tabular}{c | cc | cc | cc}
		\hline
		$n\,[\mathrm{cm}^{-3}]$
		& \multicolumn{2}{c|}{$f_{\rm cross}\,{\rm [Gyr^{-1}]}$}
		& \multicolumn{2}{c|}{$\tau_{\rm cross}\,{\rm [kyr]}$}
		& \multicolumn{2}{c}{$\eta_{\rm res}$} \\
		\cline{2-3} \cline{4-5} \cline{6-7}
		& L20 & E24 & L20 & E24 & L20 & E24 \\
		\hline
        $\geq$1 & 1198 & 627 & 64 & 142 & 0.18 & 0.47 \\
        $\geq$3 & 540 & 539 & 59 & 138 & 0.06 & 0.26 \\
        $\geq$10 & 156 & 299 & 54 & 138 & 0.01 & 0.1 \\
        $\geq$30 & 36 & 112 & 51 & 135 & 0.002 & 0.03 \\
        $\geq$100 & 5 & 15 & 46 & 124 & 0.0002 & 0.002 \\
        $\geq$200 & 1 & 2 & 44 & 113 & 0.0001 & 0.0003 \\
        $\geq$300 & 0.51 & 0.6 & 44 & 110 & 2e-05 & 0.0001 \\
		\hline
	\end{tabular}
\end{table}

\begin{figure}
    \centering
    \includegraphics[width=0.49\textwidth]{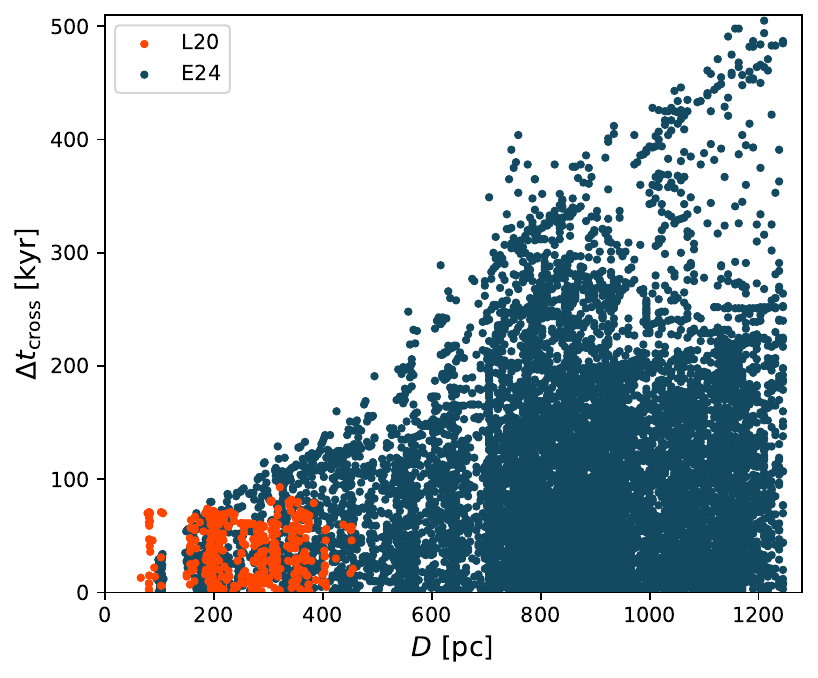}
    \caption{Duration of individual crossings as a function of distance from the map center, shown for an example density threshold of $5\,{\rm cm}^{-3}$. For the E24 map, the increase in crossing duration with distance reflects the correlation between radial distance and cell size.}  
    \label{fig:time_distance_correlation}
\end{figure}

\begin{figure*}
    \centering
    \includegraphics[width=0.95\textwidth]{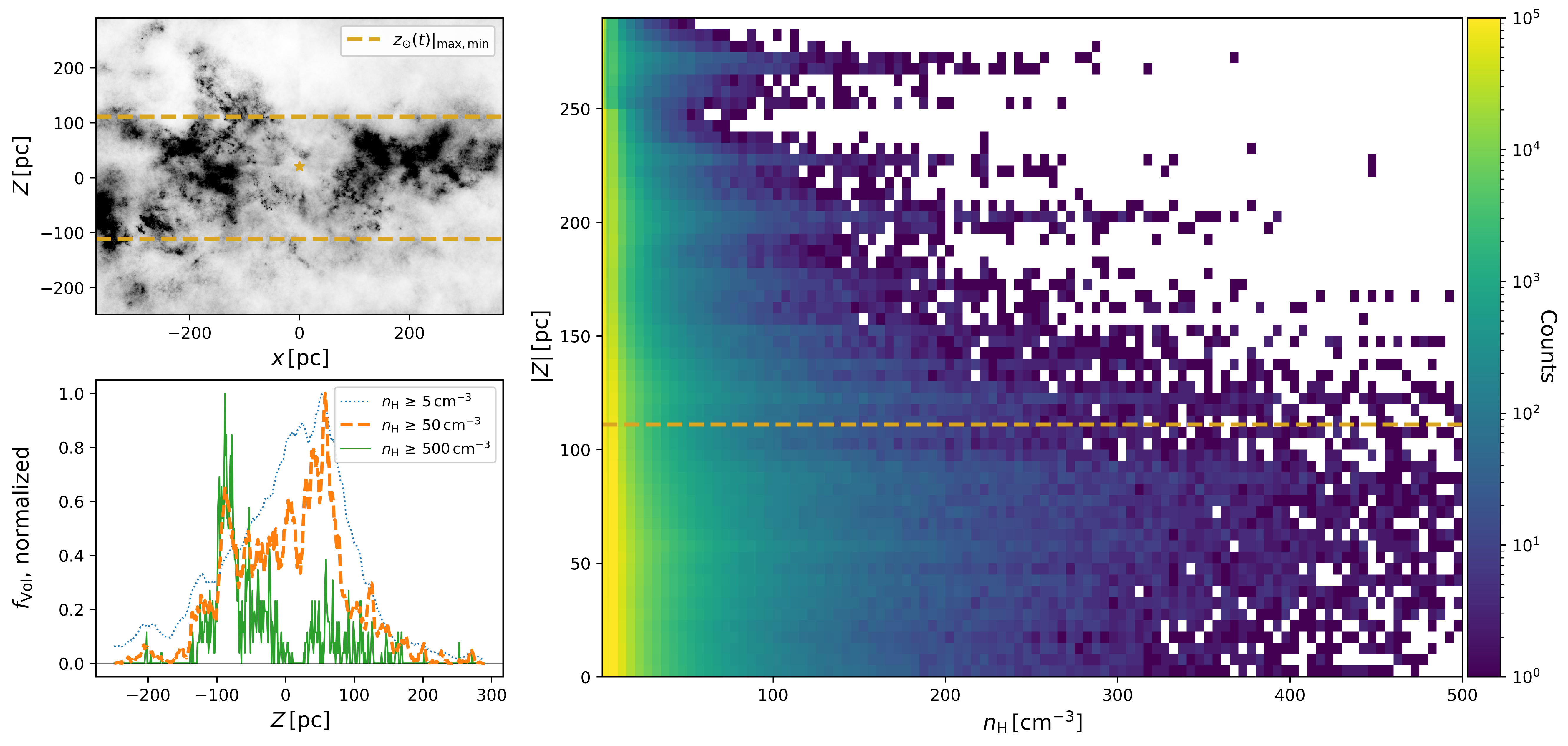}
    \caption{Vertical stratification of the ISM in the L20 map. 
    \textit{Upper left}: side projection of the L20 map. The horizontal lines mark the range of vertical oscillation ($z_\odot(t)$) currently sampled by the Solar System. 
    \textit{Lower left}: normalized volume filling factor, $f_{\rm{Vol}}$, as a function of distance from the Galactic midplane and for different density thresholds. 
    \textit{Right}: $z$-distribution of the map density bins.}  
    \label{fig:L20_zStratification}
\end{figure*}

\begin{figure*}
    \centering
    \includegraphics[width=0.95\textwidth]{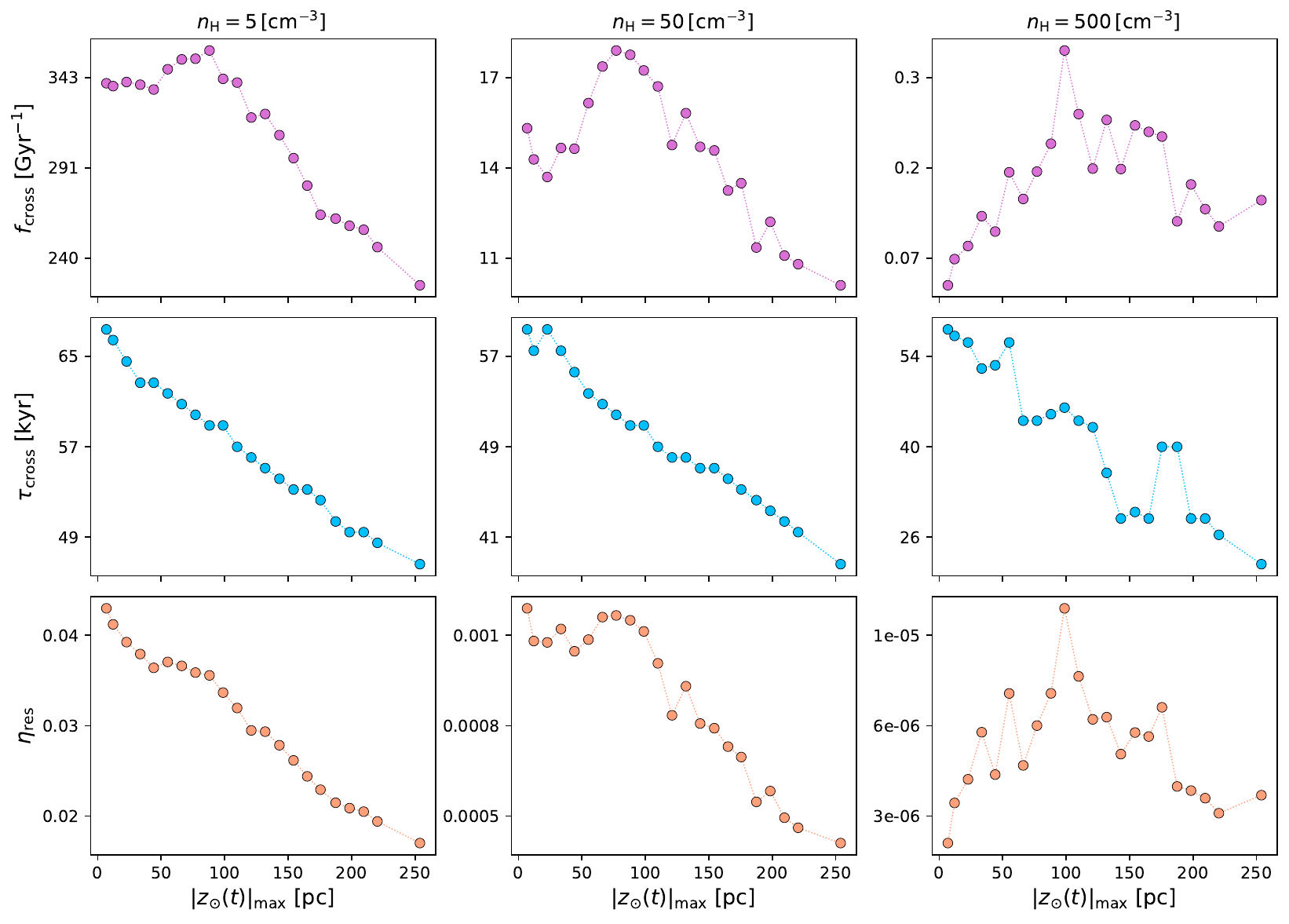}
    \caption{Encounter statistics as a function of the maximum vertical height reached by the test particle in the L20 map. The particles are initialized with Solar System-like phase-space coordinates, but placed at $z=0$ and assigned modified vertical velocities in order to sample different maximum heights.
    Each column corresponds to a different density threshold, $n_{\rm H}=5$, $50$, and $500\,{\rm cm}^{-3}$. 
    From top to bottom, the rows show the crossing frequency, the median duration of individual crossings, and the residence time fraction.} 
    \label{fig:integrationHeightsTest}
\end{figure*}

\begin{figure*}
    \centering
    \includegraphics[width=0.95\textwidth]{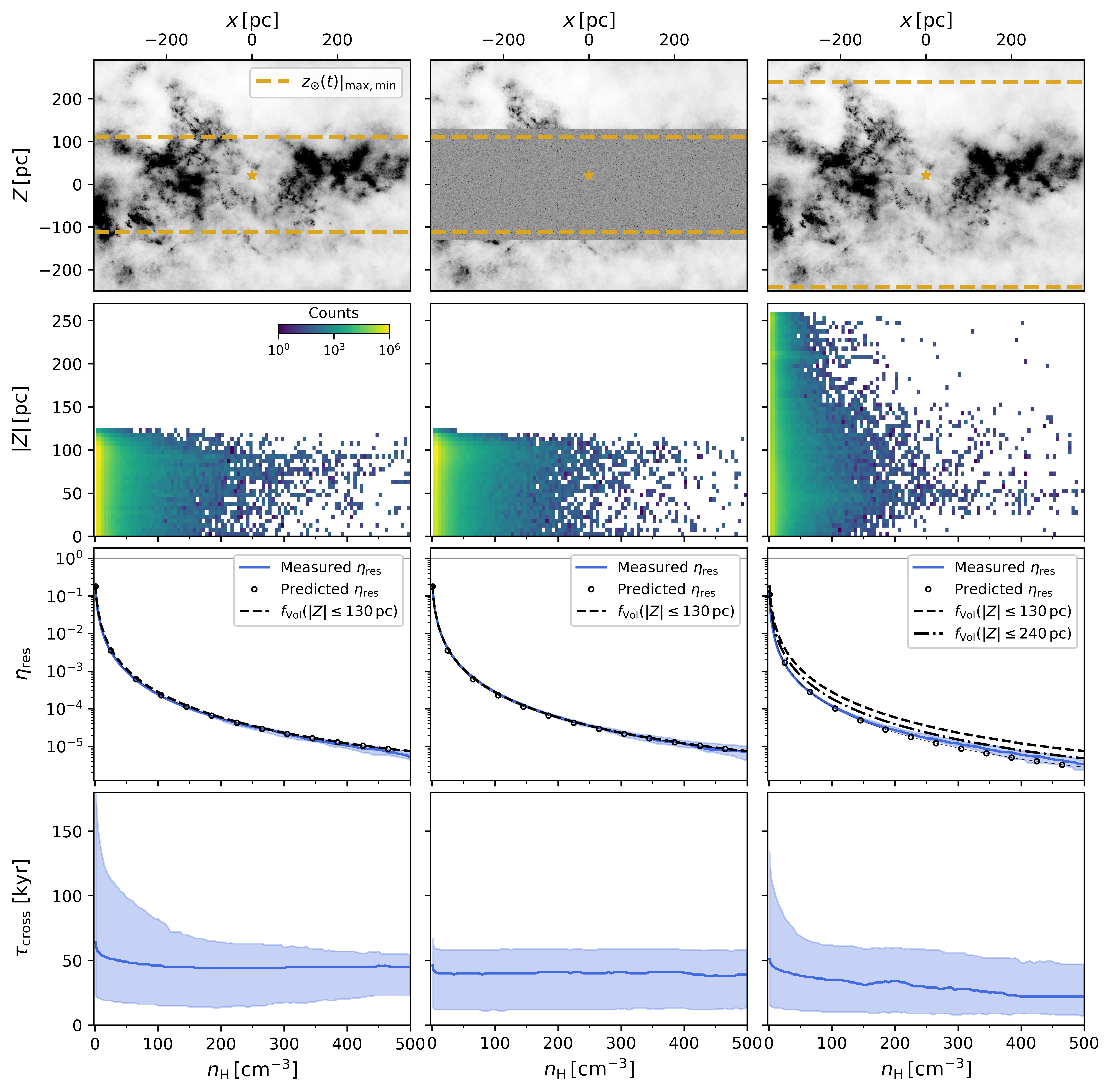}
    \caption{Crossing statistics under different realizations of the ISM density field and orbit integration. Each column corresponds to a different experiment. 
    \textit{Left column}: fiducial simulation. 
    \textit{Middle column}: identical setup, but with the ISM in the L20 map randomly shuffled within $|z|\leq130$ pc, preserving the gas distribution while removing spatial correlations. 
    \textit{Right column}: simulation with initial conditions chosen such that the Sun reaches heights where the vertical stratification of the disk changes significantly. 
    \textit{From top to bottom}: projected L20 density map, with the horizontal lines marking the range of vertical oscillation ($z_\odot(t)$) sampled by the Solar System along its orbit; 
    $z$-distribution of the density bins encountered along the orbit (for all the orbit realizations performed); 
    measured residence time fraction (solid line), compared with the predicted one (dotted line) and the volume filling factor of the sampled region (dashed line); 
    median cloud crossing time as a function of density.}  
    \label{fig:integrationTests_etaPredicted}
\end{figure*}

\section{Results and Discussion}\label{Sect:results}

\subsection{Encounter statistics}\label{sect:results-crossStats}

The Solar System encounters a variety of Galactic environments along its orbit, spanning a broad range of gas densities. In Fig.~\ref{fig:setup_overview}, we show an example of the density crossed by the Sun along its trajectory obtained by using the set-up described in Sect.~\ref{sect:methods-orbitInt}. The trajectory is characterized by intermittent passages through higher-density regions (typically of order $\sim10 - 10^2$~cm$^{-3}$), separated by longer intervals spent in lower-density environments.

We quantified this encounter history using the statistics defined in Sect.~\ref{sect:methods-crossStats}. The results are shown in Fig.~\ref{fig:crossing_statistics} for the reference integration of both the L20 and E24 maps. In Table~\ref{tab:crossing_statistics}, we report the corresponding central values at representative density thresholds.

We find that the crossing frequency, $f_{\rm cross}$, decreases smoothly with increasing density threshold. 
For densities higher than $n \geq 1 \,\rm{cm}^{-3}$, the inferred crossing frequency is greater than $600\,\rm{Gyr}^{-1}$ for both maps, corresponding to one crossing every $\sim0.8-1.6$~Myr. For $n \geq 10 \,\rm{cm}^{-3}$, the inferred crossing frequency is $\sim155\,\rm{Gyr}^{-1}$ using L20 and $\sim300\,\rm{Gyr}^{-1}$ using E24, corresponding to one such crossing every $\sim6$~Myr for L20 and $\sim3$~Myr for E24, respectively.
Crossings of very dense regions are rare: at $n \sim 300\,\rm{cm}^{-3}$, the inferred frequency decreases to $\sim 0.5\,\rm{Gyr}^{-1}$. Nevertheless, both maps predict at least one crossing per Gyr for thresholds around $200\,\rm{cm}^{-3}$.

A direct comparison with previous estimates of $f_{\rm cross}$ in the literature is not straightforward, because earlier studies generally count encounters with assumed cloud populations rather than defining contiguous crossings above a fixed density threshold, as done here. 
In an early estimate based on a simplified population of interstellar clouds, \citet{Talbot1977} found that the Sun should encounter around $31\,\rm{Gyr}^{-1}$ clouds with $n_{\rm H}\gtrsim100\,{\rm cm}^{-3}$ and about $4\,\rm{Gyr}^{-1}$ clouds with $n_{\rm H}\gtrsim10^3\,{\rm cm}^{-3}$. 
More recent estimates have focused mainly on encounters with giant molecular clouds (GMCs). \citet{Kokaia2019} modeled the Galactic distribution of GMCs using a spiral arm pattern based on \citet{Hou2014} and found an average encounter rate of $1.6\pm1.3\,{\rm Gyr}^{-1}$ for the Sun. Similarly, \citet{Moreira2025} found that Sun-like particles encounter a GMC approximately every $700\,{\rm Myr}$. These values are comparable to our crossing frequencies only if the relevant GMC densities correspond to thresholds of order $n_{\rm H}\sim100\,{\rm cm}^{-3}$. For this reason, we interpret the agreement with GMC-based studies as qualitative rather than one-to-one.

The estimated median duration of individual crossings, $\tau_{\mathrm{cross}}$, is of the order of several tens to several hundreds of kyr over a broad range of density thresholds.
For the L20 map, the median duration of crossings above $1\,\rm{cm}^{-3}$ is  $\sim64^{+140}_{-40}\,\rm{kyr}$; this decreases to $\sim54^{+80}_{-34}\,\rm{kyr}$ for $n\geq10\,\rm{cm}^{-3}$. At higher density, $\tau_{\mathrm{cross}}$ becomes nearly constant, with a median values around $45\,\rm{kyr}$.
For the E24 map, the median duration of crossings above $1\,\rm{cm}^{-3}$ is  $\sim142^{+530}_{-100}\,\rm{kyr}$. At higher densities, the median duration remains approximately constant and reaches $\sim124^{+140}_{-90}\,\rm{kyr}$ at $100\,\rm{cm}^{-3}$, with little variation for higher thresholds.

This flattening should not be interpreted as a physical saturation of cloud sizes, but rather as a consequence of the finite spatial resolution of the maps. In principle, crossings above higher density thresholds are expected to become shorter, because denser structures occupy progressively smaller volumes. In the reconstructed L20 and E24 density fields, however, the highest density structures are not fully resolved. For the L20 map, which is defined on a uniform Cartesian grid, it is significant that the inferred crossing duration approaches the characteristic time required for the Sun to cross a $1\,\rm{pc}$ resolution element at the average relative velocity between the Sun and the map, as shown in Fig.~\ref{fig:crossing_statistics}.
The E24 map is additionally affected by its native spatial discretization. E24 uses a set of concentric spherical shells, logarithmically spaced in radius and with a HEALPix pixelization (see Sect.~\ref{Sect.:Data}). The volume of the spatial bins therefore increases with distance from the map center, which affects the inferred crossing duration. Figure~\ref{fig:time_distance_correlation} illustrates this effect by showing the time spent inside individual bins as a function of distance from the map center. 

For the residence time fraction, $\eta_{\mathrm{res}}$, the experiment with the L20 map shows that the Sun spends about $18\%$ of its time in regions with densities above $1\,\rm{cm}^{-3}$, and about $1\%$ of its time above $10\,\rm{cm}^{-3}$. The residence fraction decreases further at higher thresholds, consistent with the fact that dense gas occupies a progressively smaller volume fraction.
For E24, the corresponding residence fractions are larger: about $47\%$ of the time is spent in regions above $1\,\rm{cm}^{-3}$, and about $10\%$ of its time above $10\,\rm{cm}^{-3}$. 
These values are also affected by the bin volume limitation described above.
These results imply that in the reconstructed density field the Sun spends most of its time in low density environments.
In particular, using L20 as reference because of its uniform spatial grid, the Sun remains below $10\,\rm{cm}^{-3}$ for approximately $99\%$ of the sampled time, and below $1\,\rm{cm}^{-3}$ for approximately $80\%$ of the time.
Although this is a first order estimate, it is consistent with the presence of several hundred-parsec-sized voids in the solar neighborhood, as observed from the top down view of the reconstructed maps (see Fig. \ref{fig:maps_overview}), and as similarly seen in nearby galaxies \citep[see e.g.,][]{Lee2023,Barnes2023}.

A related estimate for $\eta_{\mathrm{res}}$ was obtained by \citet{Pfalzner2020}, who studied the fraction of time spent by interstellar objects inside GMCs using the simulation set-up of \citet{Kokaia2019}. They found that typical interstellar objects in the Solar neighborhood, with velocities of order $30\,{\rm km\,s^{-1}}$, spend only about $0.1$ to $0.2\%$ of their time inside GMCs, while slower ISOs, with velocities of order $5\,{\rm km\,s^{-1}}$, spend about $1$ to $2\%$ of their time inside GMCs. Although these values are not directly comparable to our residence fractions, because they refer to ISO trajectories through an idealized GMC population rather than to the Solar orbit through reconstructed density fields, they provide a useful reference point for interpreting residence fractions in dense environments.

In general, the derived crossing statistics are subject to several limitations. 
First, high density structures are not fully captured by the maps (see Sect.~\ref{Sect.:Data}). 
However, because such regions occupy a small volume fraction, their contribution is expected to be limited, although they may dominate the statistics of rare high density encounters. 
Second, the results are affected by the native spatial discretization of the maps, in particular for E24, as discussed above. 
Third, the effective resolution and smoothing of the maps can dilute unresolved dense structures below their peak densities, leading to an underestimate of the crossing statistics at high density thresholds. Smoothing can also broaden density structures, affecting the inferred crossing durations and residence fractions at lower thresholds.
Finally, the intrinsic motions of the ISM are not captured by these maps. Gas motions could modify the true duration of an encounter by changing the relative velocity between the Sun and the gas, either shortening or lengthening individual crossings depending on the direction of the motion. We therefore interpret these quantities as estimates derived from the reconstructed density fields, rather than as direct measurements of the true physical ISM values.

Despite these limitations, the crossing statistics provide useful insight into the encounter history of the Solar System. It is notable that the two maps yield broadly similar trends despite their different spatial extents. The L20 map reconstructs the nearby $\sim400\,\rm{pc}$ region around the Sun, whereas E24 extends to $\sim1.25\,\rm{kpc}$ and includes larger Galactic structures \citep[see e.g.,][]{Kormann2026}. 
This similarity suggests that the central values of the encounter statistics are primarily sensitive to the density distribution sampled by the orbit, rather than to the detailed large scale organization of the ISM. This interpretation is also supported by a test in which we repeated the calculations using a random orientation of the map, which gives similar results (Appendix~\ref{sect:ValidationTests}). We investigate this point further in Sect.~\ref{sect:results-ISM_stratification}, where we study how the ISM structure and the Solar orbit affect these quantities.
In Appendix~\ref{sect:ValidationTests}, we also show that the encounter statistics are robust to changes in the Solar initial conditions and that restricting E24 to the L20 volume improves the agreement between the two maps.

\subsection{Vertical ISM stratification and the solar orbit} \label{sect:results-ISM_stratification}

In this section, we investigate how the encounter statistics are affected by the vertical stratification of the ISM and by the Solar orbit, focusing in particular on the Sun's vertical motion. In details, we aim to identify the density regimes and ranges of vertical heights sampled by the Sun in which the statistics are mainly controlled by the gas density distribution itself, and those in which the coupling between the ISM structure and the Solar motion becomes important. For this part of the analysis, we use the L20 map, taking advantage of its uniform Cartesian grid.

The different encounter statistics are sensitive to different properties of the gas density and the Sun's orbit. The crossing frequency and crossing duration depend not only on the gas distribution, which sets the probability of encountering gas above a given threshold, but also on the spatial organization of the gas. A density field in which dense gas is concentrated into a small number of coherent structures produces fewer and longer crossings than one in which the same dense gas fraction is distributed among many smaller structures. Both quantities depend on the relative velocity between the Sun and the map, since larger relative velocities shorten individual crossings and can increase the number of structures crossed per unit time.
By contrast, the residence fraction is less sensitive to the detailed spatial organization of the gas, as it measures the total fraction of time spent above a given density threshold, rather than the number or duration of individual crossings. All the quantities are sensitive to the heights sampled by the Sun and the gas stratification of the disk.

To investigate this dependence, we first examine the vertical structure of the ISM in the L20 map. Figure~\ref{fig:L20_zStratification} shows this analysis through a set of diagnostic plots. 
The side projection of the map (upper left panel) shows that the gas is organized into structures which appear to be uniformly distributed over a vertical range, roughly within $|z|\,\lesssim 100 -130\,{\rm pc}$, rather than being concentrated around $z=0 \, {\rm pc}$. 
This behavior is also visible in the normalized volume filling factor, $f_{\rm Vol}$, shown in the lower left panel for representative low, intermediate, and high density thresholds ($n_{\rm H}\geq 5 ,\,50,\,500\,{\rm cm}^{-3}$). For the lowest threshold, the filling factor is largest near the Galactic midplane and decreases gradually with increasing $|z|$, broadly consistent with the expected vertical stratification of Galactic dust \citep[see e.g.,][]{Misiriotis2006,Guo2025}. At higher density thresholds, however, dense structures appear to populate more uniformly the vertical heights before declining rapidly at heights with $|z|\,\geq100-130 \, {\rm pc}$. The same trend is reflected in the vertical distribution of the density bins shown in the right panel.
This behavior may partially be explained by the limited volume sampled by the L20 map, rather than reflect the global vertical structure of the Galactic ISM, but it might also suggest that different density regimes may have different effective vertical distributions. Low density gas may be primarily governed by gravity, while denser structures may be shaped by feedback processes.
It is worth noting that a similar behavior is also found when using the larger E24 map, although in this case the interpretation is affected by the bin size distance dependence of the native grid. We briefly discuss this in Appendix~\ref{sect:E24_stratification}. However, a detailed investigation of the scale heights dependency on density thresholds is beyond the scope of this work. 

To test the role of the Solar vertical motion, we modified the initial $z$ and $v_z$ of the test particle, while keeping the remaining phase space coordinates fixed to the reference values.
The resulting encounter statistics are shown in Fig.~\ref{fig:integrationHeightsTest} for representative density thresholds. 
This experiment isolates how the statistics change when the orbit samples different portions of the vertically stratified ISM.

Among the three quantities considered, $\tau_{\rm cross}$ shows the clearest dependence on the maximum height sampled by the Sun: it decreases with increasing vertical amplitude for all density thresholds. This trend can be explained by the fact that orbits reaching larger $|z|$ have larger vertical velocities when traversing the disk. As discussed in Sect.~\ref{sect:results-crossStats}, $\tau_{\rm cross}$ approaches the resolution limit of the map, corresponding to the time required to cross a spatial resolution element. Because of this limited resolution, we cannot investigate in detail whether the sampling angle of the orbit might affect the inferred crossing durations.

The crossing frequency and the residence time show a more complex behavior, which depends on the adopted density threshold. Since these statistics are approximately related by relation~\ref{eq:crossStatistics-relation}, their trends are not completely independent. 
At low density thresholds, $f_{\rm cross}$ remains approximately constant as long as the Sun samples heights within $|z|\lesssim100\,{\rm pc}$, and then appears to decline at larger vertical amplitudes. Instead, $\eta_{\rm res}$ decreases more gradually.
At intermediate density thresholds, corresponding to regimes where the vertical stratification of the ISM begins to deviate from a simple monotonic decline, both $f_{\rm cross}$ and $\eta_{\rm res}$ remain relatively stable while the Sun samples the region within $|z|\lesssim100 -130\,{\rm pc}$. A clearer decrease appears only when the orbit reaches heights beyond this range. 
At the highest density thresholds, the statistics become more sensitive to the particular locations of dense structures within the finite volume of the L20 map. In this regime, neither the crossing frequency nor the residence fraction shows a simple monotonic decrease with maximum vertical height.

These results suggest that, within the local volume sampled by the L20 map, the probability of encountering dense ISM structures does not necessarily decrease smoothly with distance from the Galactic midplane, as found in other studies \citep[see e.g.,][]{Kokaia2019}. 
However, in \citet{Kokaia2019}, the vertical decrease follows directly from the assumed distribution of GMCs, and the sampled range of heights extends well beyond the one considered here. When we consider orbits reaching larger vertical amplitudes ($|z|\geq130\,\rm{pc}$), we also find that the encounter probability decreases. 
Our findings also suggest that the Solar vertical oscillation through the Galactic midplane does not necessarily translate into a periodic perturbation of the Solar System by dense ISM structures, as sometimes invoked in relation to periodicities in the terrestrial impact cratering record \citep[see e.g.,][]{Rampino2015}.
The result that dense structures do not appear to follow a simple monotonic decline with height may be also relevant for studies of stellar cluster dissolution. For example, \citet{Moreira2025} successfully modeled the open cluster scale height using vertically decreasing encounter probabilities with molecular clouds. It would therefore be interesting to explore whether alternative assumptions, for example a more uniform probability within a finite vertical range (e.g., a sigmoid-like vertical dependence), affect the inferred evolution of open cluster populations.

The residence fraction provides the clearest way to isolate the coupling between the vertical structure of the ISM and the Solar orbit. Unlike the crossing frequency and crossing duration, it does not depend directly on how the gas is spatially organized. Instead, it measures only the total fraction of time spent above a density threshold. If this is correct, the residence fraction can be approximated as
\begin{equation} \label{eq:eta_predicted}
    \eta_{\rm res}^{\rm predicted}(\geq n_{\rm th}) = \int_{z_{\rm min}}^{z_{\rm max}} p_{\odot}(z) \, f_{\rm Vol}(\geq n_{\rm th}|z) \, {\rm d}z,
\end{equation}
where $p_{\odot}(z)$ is the fraction of time spent at a given $z$, $f_{\rm Vol}(\geq n_{\rm th}\mid z)$ is the volume filling factor of gas above the threshold at $z$, and $z_{\rm min}$ and $z_{\rm max}$ are the minimum and maximum heights sampled by the Solar orbit.
If $f_{\rm Vol}$ is approximately independent of $z$ over the vertical range sampled by the Sun, this expression reduces to
\begin{equation}\label{eq:eta_predicted_simpl}
    \eta_{\rm res}^{\rm predicted}(\geq n_{\rm th}) = f_{\rm Vol}(n\geq n_{\rm th}| z_{\rm min}\leq z \leq z_{\rm max}).
\end{equation}
In this limiting case, the residence fraction is determined by the gas distribution within the sampled region, rather than by the detailed vertical coupling between the orbit and the ISM.

We test this interpretation using the fiducial integration set-up and two additional experiments. 
In the first experiment, we randomly shuffle the L20 density field within $|z|\leq130\,{\rm pc}$. This preserves the gas distribution within the vertical range sampled by the Solar orbit, while removing the spatial correlations of the gas. 
In the second experiment, we use initial conditions for which the Sun reaches larger heights, where the vertical stratification of the disk changes more significantly. 
The results are presented in Fig.~\ref{fig:integrationTests_etaPredicted}.

We find that Eq.~\ref{eq:eta_predicted} successfully predicts the residence fraction in all three cases. In the fiducial and shuffled experiments, the residence fraction is also well approximated by the volume filling factor computed for the sampled vertical range, thus satisfying the approximation from Eq.~\ref{eq:eta_predicted_simpl}.  This indicates that, for the present Solar orbit, the $\eta_{\rm res}$ is mainly controlled by the gas distribution.
In the shuffled experiment, however, $\tau_{\rm cross}$ becomes nearly constant, reflecting the loss of coherent density structures.
In the experiment with a larger vertical amplitude, the volume filling factor computed over the sampled region is no longer sufficient to reproduce $\eta_{\rm res}$. In this case, the coupling between the Solar vertical motion and the vertical stratification of the ISM becomes important.

These tests show that the residence fraction is primarily controlled by the combination of the vertical volume filling factor and the fraction of time spent by the Sun at each height. By contrast, the crossing duration and crossing frequency depend more strongly on the spatial organization of the gas, because they are sensitive to how structures are organized. 

\subsection{Interstellar gas flux and dust exposure} \label{sect:results-gasFlux}

\begin{figure}
    \centering
    \includegraphics[width=0.49\textwidth]{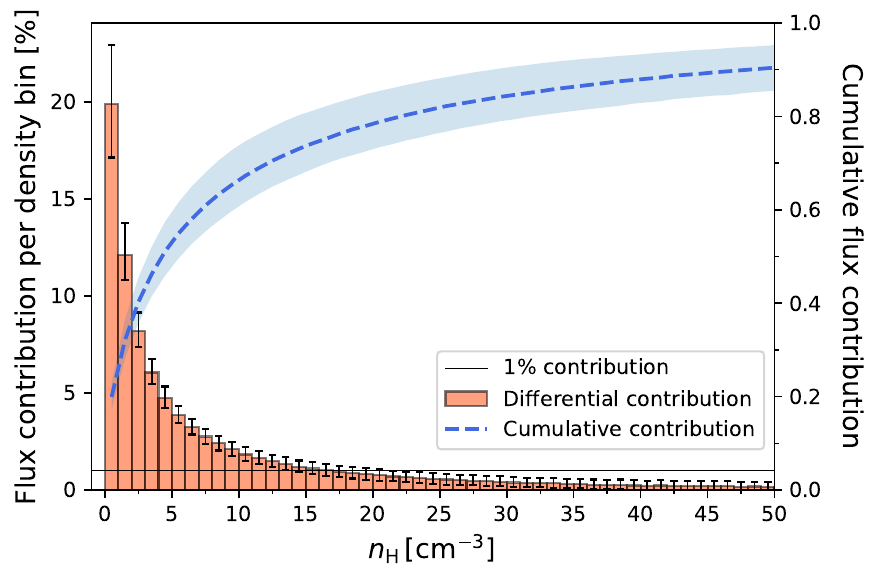}
    \caption{Contribution to the time averaged gas flux as a function of gas density. The histogram shows the differential contribution from individual density bins, while the dashed blue line shows the cumulative contribution from gas below a given density threshold. The horizontal black line marks a contribution of 1\%. We show density bins between 0 and 50$\,{\rm cm}^{-3}$, with a bin width of $1\,{\rm cm}^{-3}$.}  
    \label{fig:flux_diff_cum_contribution}
\end{figure}

We next estimate the amount of interstellar material encountered by the Solar System along its orbit. This quantity depends not only on the gas density sampled along the trajectory, but also on the relative velocity between the Sun and the surrounding ISM. We therefore compute the time averaged gas flux, $\Phi_{\rm gas}^{\rm tot}$, using Eq.~\ref{eq:mean_gas_flux}, and quantify both the contribution from individual density ranges and the corresponding accreted interstellar dust mass. The estimates here presented are derived from the L20 map.

We obtain a time averaged gas flux along the sampled Solar orbit of $\Phi_{\rm gas}^{\rm tot} = (6.0 \pm 0.9) \cdot 10^{13} \, [\rm{cm^{-2} \cdot yr^{-1}}]$.
However, the contribution to this flux is not uniform across density ranges and in Fig.~\ref{fig:flux_diff_cum_contribution} we report the differential and cumulative contribution to $\Phi_{\rm gas}^{\rm tot}$ by density. 
Most of the contribution to the flux is coming from relatively low density environments: 50\% of the total flux comes from gas with $n_{\rm H}\lesssim3-4\,{\rm cm}^{-3}$, while 90\% is produced by gas with $n_{\rm H}\lesssim48\,{\rm cm}^{-3}$. Gas in the lowest density bin, $0<n_{\rm H}\leq1\,{\rm cm}^{-3}$, already accounts for about 20\% of the total flux. This contribution should be interpreted with caution as such low density environments are among the most difficult regimes to reconstruct accurately with dust extinction maps. 
As a simple test, we added a random isotropic velocity component to the gas. Each Cartesian component was drawn from a Gaussian distribution with zero mean and dispersion $\sigma_{\rm ISM}=8.9\,{\rm km\,s^{-1}}$ \citep{Marasco2017}. This increases the total flux slightly, to $\Phi_{\rm gas}^{\rm tot}\simeq 7.3\times10^{13}\,{\rm cm^{-2}\,yr^{-1}}$, while leaving the relative contribution of different density ranges essentially unchanged. Since a fully consistent treatment of gas motions would also modify the cloud crossing times, we use the static map calculation as our reference case and interpret this test only as an estimate of the possible effect on adding a random isotropic velocity.

The undisturbed ISM flow outside the heliosphere carries both gas and dust, but it is strongly modified as it interacts with the Solar wind. In particular, the transport of interstellar dust into the Solar System depends also on grain size, charge, and heliospheric conditions \citep[see e.g.,][]{Sterken2015,Kruger2015}. Nevertheless, a fraction of interstellar dust grains can enter the Solar System and the Earth's atmosphere \citep[see e.g.,][]{Grun1993,Taylor1996,Altobelli2005}. 
We therefore convert the time averaged gas flux into an estimate of the accreted interstellar dust mass under the simplifying assumptions described in Sect.~\ref{sect:methods-gasFlux} and using Eq.~\ref{eq:mass_rate_full_eq_simplyfied}. 
Using the reported parameters values, we obtain
\begin{equation}
\dot M_{\rm dust} \approx  \, \sigma_{\rm \oplus} \, \epsilon \,\, 1.4 \cdot 10^{-15} \, {\rm [kg \cdot  cm^{-2} \cdot yr^{-1}]}.
\end{equation}
This value represents the orbit averaged estimate over the sampled Solar trajectory. However, the Solar System is currently moving through a low density ISM environment, known as the CLIC \citep[see \S \ref{sect:intro} and, e.g.,][]{Redfield2008}. For comparison with present day conditions, we therefore also compute the dust mass accretion rate using only the flux measured for $n_{\rm H}<1\,{\rm cm}^{-3}$. In this case, we find
\begin{equation}
\dot M_{\rm dust} \approx  \, \sigma_{\rm \oplus} \, \epsilon \,\,3.4 \cdot 10^{-16} \, {\rm [kg \cdot  cm^{-2} \cdot yr^{-1}]},
\end{equation}
which corresponds to $\dot M_{\rm dust} \approx \, \epsilon \,430 \, {\rm [kg \cdot yr^{-1}]}.$
Matching a present day interstellar dust accretion rate on Earth of order $100\,{\rm kg\,yr^{-1}}$ \citep[see e.g.,][]{Strub2019} would therefore require an effective delivery efficiency of $\epsilon\approx0.23$. As discussed in Sect.~\ref{sect:methods-gasFlux}, $\epsilon$ absorbs unresolved focusing, heliospheric filtering, and grain size and charge dependent transport effects and should therefore be viewed as a very first order estimate.

Finally, we examined how the incident flux and accreted dust mass vary with distance from the Galactic midplane. The incident flux reflects the combined effect of the density distribution and the relative velocity along the orbit. As shown in the left panel of Fig.~\ref{fig:flux_massAccreation_zDependency}, the mean instantaneous flux is slightly enhanced near the midplane, where the vertical velocity is larger.
The fractional contribution to the total accreted dust mass depends additionally on the amount of time spent by the Sun at each height. As a result, the heights associated with the largest instantaneous flux are not necessarily those that dominate the total accreted mass over the full orbit. This is illustrated in the right panel of Fig.~\ref{fig:flux_massAccreation_zDependency}, where the largest contribution comes from heights close to the vertical turning points of the Solar orbit.

\subsection{Caveats}

The encounter statistics presented in this work are subject to several limitations:
\begin{itemize}
    \item the maps represent local, present day, static conditions ISM;
    \item the formation, evolution, and dispersal of clouds are neglected;
    \item the maps have finite spatial resolution and do not resolve compact high density regions (see Sect.~\ref{Sect.:Data});
    \item the spatial grid on which each map is defined can influence the inferred encounter statistics, especially the crossing durations (see Sect.~\ref{sect:results-crossStats});
    \item intrinsic gas motions and transient events, such as SNe, are not included.
\end{itemize}

Despite these limitations, the analysis provides a useful first order, data driven estimate of the density environments encountered by the Sun and of the corresponding gas and dust exposure along its Galactic orbit. The results should therefore be interpreted as observationally informed estimates based on the reconstructed local ISM, rather than as an accurate reconstruction of the Solar System's actual encounter history.


\section{Conclusions}\label{Sect.:conclusions}

In this work, we used 3D dust maps of the Solar neighborhood to characterize the interstellar environments encountered by the Solar System along its Galactic orbit. We integrated the Solar orbit through the reconstructed density fields and measured the frequency, duration, and residence time fraction of passages above different gas density thresholds. We also estimated the corresponding interstellar gas flux and dust exposure.
Our main results can be summarized as follows:

\begin{itemize}

\item The Solar System samples a broad range of interstellar densities along its orbit, but spends most of its time in low density environments.
The crossing frequency decreases rapidly with increasing density threshold. For $n_{\rm H}\geq1\,{\rm cm}^{-3}$, both maps give $f_{\rm cross}>600\,{\rm Gyr}^{-1}$. For $n_{\rm H}\geq10\,{\rm cm}^{-3}$, the frequency decreases to $\sim155\,{\rm Gyr}^{-1}$ in L20 and $\sim300\,{\rm Gyr}^{-1}$ in E24. Crossings of very dense gas are rare, with $f_{\rm cross}\sim0.5\,{\rm Gyr}^{-1}$ at $n_{\rm H}\geq300\,{\rm cm}^{-3}$, although both maps predict one crossing per Gyr for thresholds around $200\,{\rm cm}^{-3}$. 
The median duration of individual crossings is typically several tens of kyr, for example $\sim64\,{\rm kyr}$ in L20 and $\sim142\,{\rm kyr}$ in E24 for $n_{\rm H}\geq1\,{\rm cm}^{-3}$. At higher density thresholds, these durations become resolution limited and should be interpreted as likely upper limits on the true crossing times of unresolved compact structures. 
For the residence fraction, we found that the Sun spends $\sim18\%$ of the sampled time above $1\,{\rm cm}^{-3}$ in L20 and $\sim47\%$ in E24, while the corresponding fractions above $10\,{\rm cm}^{-3}$ are about $1\%$ and $10\%$. 
The numerical differences between L20 and E24 reflect their different spatial discretizations, with E24 being more affected as it is defined on a spherical grid.
\item The vertical gas stratification in the maps depends on the density threshold considered. Low density gas is more likely to be found near the Galactic midplane, with its volume filling factor decreasing gradually at larger $|z|$. Higher density structures, instead, appear to populate more uniformly the vertical heights before declining rapidly at heights with $|z|\geq 100  -  130\, \mathrm{pc}$.
Consistently with this, the crossing frequency of dense structure remains relatively stable for Solar orbits confined within this vertical range. This suggests that Galactic midplane passages do not necessarily lead to periodic perturbations of the Solar System.
\item The residence time fraction reflects the coupling between the vertical structure of the ISM and the vertical motion of the Sun. We find that it is well reproduced by weighting the volume filling factor of gas above a given density threshold at each height by the fraction of time that the Sun spends at that height. By contrast, the crossing duration and crossing frequency also depend on the spatial organization and coherence of the gas structures along the orbit.

\item The gas and dust exposure of the Solar System is dominated by relatively low density environments. We find a time averaged gas flux of $\Phi_{\rm gas}^{\rm tot} = (6.0 \pm 0.9)\times10^{13}\,{\rm cm}^{-2}\,{\rm yr}^{-1}$. About half of this flux is contributed by gas with $n_{\rm H}\lesssim3-4\,{\rm cm}^{-3}$, and about $90\%$ comes from gas with $n_{\rm H}\lesssim48\,{\rm cm}^{-3}$. 

We also converted the gas flux into a first order estimate of the dust mass accretion rate. For comparison with present day low density conditions, and under very basic assumptions, we obtain $\dot M_{\rm dust} \approx \, \epsilon \,430 \, {\rm [kg \cdot yr^{-1}]}$. This corresponds to an effective delivery efficiency of $\epsilon\approx0.23$, to match the present day $100\,{\rm kg \cdot yr^{-1}}$.
\end{itemize}

These results provide a data driven estimate of the density environments sampled by the Solar System in the local Galactic disk. They should be interpreted as a first order approximation, as they are limited by the finite extent and resolution of the maps, the assumption of a static density field, and the absence of transient events such as nearby supernovae. Nevertheless, to our knowledge, this work provides the first estimate of these quantities based on 3D observational data.

Future work based on upcoming dust-map reconstructions and numerical simulations will allow these statistics to be better constrained. Such studies will also make it possible to characterize additional environmental quantities encountered by the Sun, including temperature and magnetic field strength. These findings could be complementary to heliospheric and geological studies aimed at understanding how changes in the ISM affect the heliosphere and may leave signatures in terrestrial records.

\begin{acknowledgements}
EM, JA, and SH were co-funded by the European Union (ERC, ISM-FLOW, 101055318). This work has made use of data from the European Space Agency (ESA) mission \textit{Gaia} \href{https://www.cosmos.esa.int/gaia}{https://www.cosmos.esa.int/gaia}, processed by the \textit{Gaia} Data Processing and Analysis Consortium (DPAC,
\href{https://www.cosmos.esa.int/web/gaia/dpac/consortium}{https://www.cosmos.esa.int/web/gaia/dpac/consortium}). Funding for the DPAC has been provided by national institutions, in particular the institutions participating in the \textit{Gaia} Multilateral Agreement.
\end{acknowledgements}

\vspace{-1cm}

%
%

\bibliographystyle{aa} 
\bibliography{references} 

\begin{appendix} 
\onecolumn
\section{From differential extinction to gas volume density}\label{Sect:Ext_to_Dens}

The maps by \citet[][hereafter L20]{Leike2020} and \citet[][hereafter E24]{Edenhofer2024} are both provided in terms of the differential extinction: L20 as the \textit{Gaia} G-band optical depth per parsec ($s_{\rm{G}}=\frac{\tau_G}{1\,\rm{pc}}$) and E24 as the dimensionless extinction ($A'_{\mathrm{ZGR23}} = \frac{dA_{\mathrm{ZGR23}}}{1\,\mathrm{pc}}$) defined in \citet[][hereafter ZGR23]{Zhang2023}. To convert the differential extinction data to gas volume density, we follow the procedure of \citet{Zucker2021}, \citet{ONeill2024}, and \citet{ONeill2026}, summarized as follows.

The gas volume of a given bin is
\begin{equation}\label{eq.:gas_number_dens}
    n_{\mathrm{H}} = \frac{\#}{dv} = \frac{N_{\mathrm{H}}\,da}{dv},
\end{equation}
where $\#$ is the number of particles, $dv$ the bin volume, $da$ is the area of the bin, and $N_{\mathrm{H}}$ is the hydrogen column density.
Assuming a constant ratio between the hydrogen column density and extinction in a given band \citep[see e.g.,][]{Draine2009}, it is possible to write 
\begin{equation}
\frac{A_{\mathrm{X}}}{N_{\mathrm{H}}}=\alpha_{\mathrm{X}}\,\mathrm{cm}^2\,\mathrm{mag},
\end{equation}
with $\alpha_{\mathrm{X}}$ a coefficient that can be derived from the curve in the work by \citet{Draine2009}. Using the \textit{Gaia} G-band extinction, with a mean wavelength of $\lambda=673\,\mathrm{nm}$ \citep{Jordi2010}, the $\alpha$ coefficient is equal to $4\times10^{-22}$.

For the L20 map, given the relationship between extinction $A_{\rm{G}}$ and optical depth ($A_{\rm{G}} = 1.086 \times \tau_{\rm{G}}$), Eq.~\ref{eq.:gas_number_dens} becomes
\begin{equation}\label{eq.:gas_number_dens_1}
    n_{\mathrm{H}} = \frac{1.086 \times \tau_{\rm{G}}}{\alpha_{\mathrm{G}}} \frac{da}{dv} = \frac{1.086}{\alpha_{\mathrm{G}}} \frac{\tau_G}{1\,\rm{pc}} = 880 \, \rm{cm^{-3}} \times s_{\rm{G}}.
\end{equation}

For the E24 map, we know that the dimensionless differential extinction $A'_{\mathrm{ZGR23}}$ can be converted to a given extinction band as, 
\begin{equation}
    A'_{\mathrm{X}} = m_{\mathrm{X}}A'_{\mathrm{ZGR23}},
\end{equation}
where $m_{\mathrm{X}}$ is a coefficient from the extinction curve by ZGR23. For the \textit{Gaia} G-band extinction, $m_{\mathrm{X}}=2.036$. 
From this, Eq.~\ref{eq.:gas_number_dens} can be written as
\begin{equation}\label{eq.:gas_number_dens_2}
    n_{\mathrm{H}} = \frac{A_{\mathrm{X}}}{\alpha_{\mathrm{X}}} \frac{da}{dv} = \frac{m_{\mathrm{X}}}{\alpha_{\mathrm{X}}} \frac{dA_{\mathrm{ZGR23}}}{1 \mathrm{pc}} \frac{da\,dr}{dv} = \frac{m_{\mathrm{X}}}{\alpha_{\mathrm{X}}} \frac{dA_{\mathrm{ZGR23}}}{1 \mathrm{pc}} = 1650\,\mathrm{cm^{-3}} \times A'_{\mathrm{ZGR23}}.
\end{equation}

Recently, \citet{ONeill2026} directly compared dust extinction with H{\sc i} emission and found that the conversion factor previously adopted by \citet{ONeill2024}, $1650\,{\rm cm}^{-3}$, is underestimated by a factor of $\sim1.8$. They recommend a conversion factor of $3002\,{\rm cm}^{-3}$, corresponding to the median of the distribution of possible values. This reflects the fact that this conversion is not described by a unique best value, but instead has a calibration uncertainty. For a more detailed discussion, we refer to \citet{ONeill2026} and their Fig.~16.
We therefore assume that the same correction applies to the L20 map and multiply the conversion factor used by \citet{Zucker2021} by a factor of $1.82$. 

The use of the updated conversion factors affects the absolute gas density and therefore leads to changes in the crossing statistics. However, the main conclusions presented in the paper are not significantly affected. For comparison, Table~\ref{tab:crossing_statistics_oldConvFact} reports the crossing statistics obtained using the pre-correction conversion factors, providing an equivalent version of Table~\ref{tab:crossing_statistics}.
The quantity obtained using this procedure represents the total volume density of hydrogen nuclei, independent of phase \((n = n_{\mathrm{H\,I}} + 2n_{\mathrm{H_2}})\).

\begin{table}
	\centering
    \renewcommand{\arraystretch}{1.5}
	\caption{Equivalent of Table~\ref{tab:crossing_statistics}, but for gas volume density maps obtained using the pre-correction conversion factors.}
	\label{tab:crossing_statistics_oldConvFact}
	\begin{tabular}{c | cc | cc | cc}
		\hline
		$n\,[\mathrm{cm}^{-3}]$
		& \multicolumn{2}{c|}{$f_{\rm cross}\,{\rm [Gyr^{-1}]}$}
		& \multicolumn{2}{c|}{$\tau_{\rm cross}\,{\rm [kyr]}$}
		& \multicolumn{2}{c}{$\eta_{\rm res}$} \\
		\cline{2-3} \cline{4-5} \cline{6-7}
		& L20 & E24 & L20 & E24 & L20 & E24 \\
		\hline
        $\geq$1 & 814 & 609 & 61 & 139 & 0.1 & 0.35 \\
        $\geq$3 & 306 & 422 & 56 & 137 & 0.03 & 0.17 \\
        $\geq$10 & 73 & 189 & 52 & 137 & 0.01 & 0.05 \\
        $\geq$30 & 14 & 48 & 48 & 132 & 0.001 & 0.01 \\
        $\geq$100 & 1 & 3 & 44 & 114 & 0.0001 & 0.0004 \\
        $\geq$200 & 0.33 & 0.3 & 45 & 107 & 1e-05 & 4e-05 \\
        $\geq$300 & 0.12 & 0.06 & 45 & 97 & 4e-06 & 1e-05 \\
		\hline
	\end{tabular}
\end{table}

\FloatBarrier

\section{Validation tests}\label{sect:ValidationTests}

In this Section, we present the validation tests for the encounter statistics presented in the main part of the paper. 
We first tested how the map orientation and the different initial conditions for the Sun's orbit influence the results.
 
\begin{figure*}
    \centering
    \includegraphics[width=0.98\textwidth]{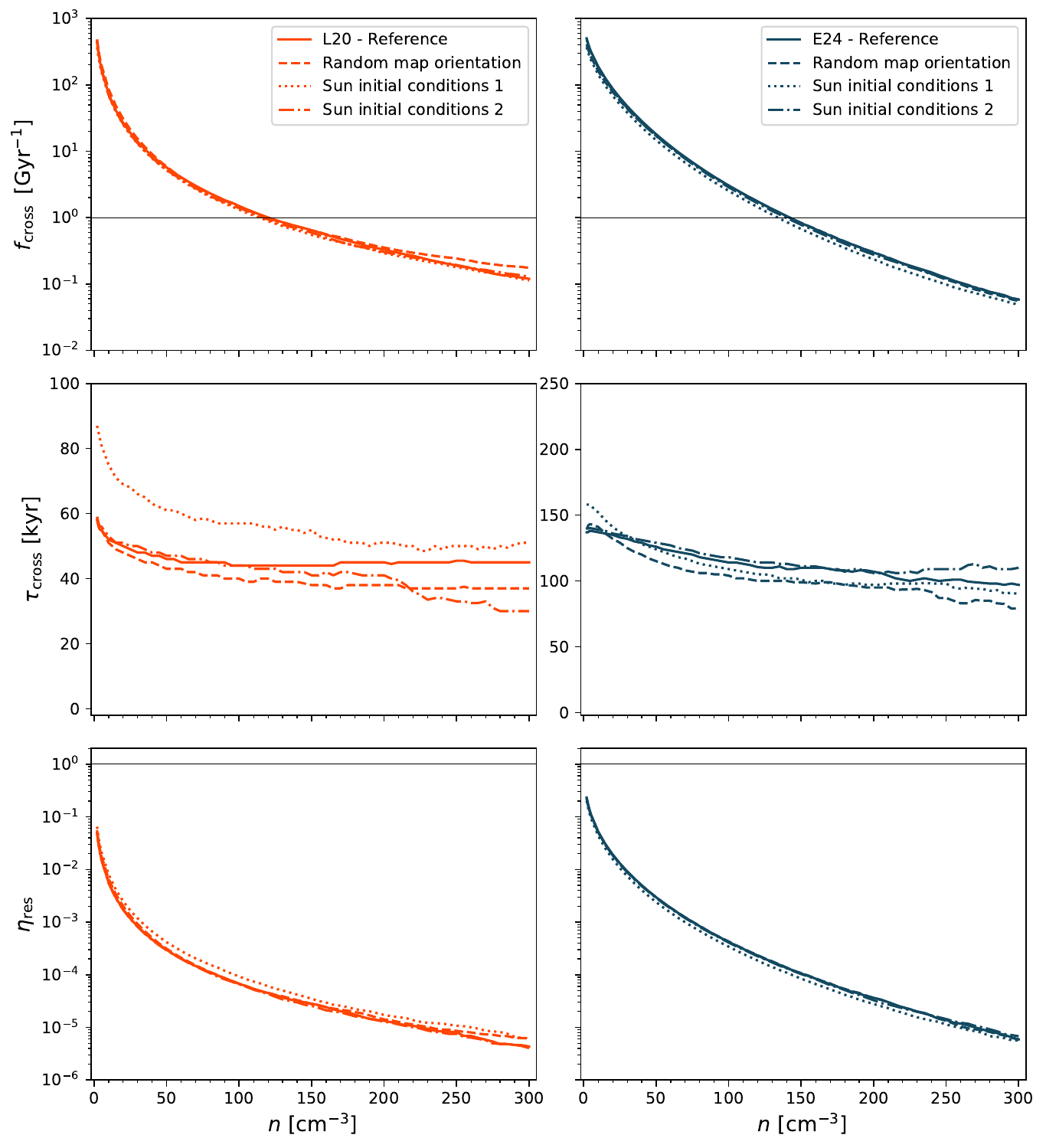}
    \caption{Different tests for the encounter statistics. The \textit{left column} shows the results for the L20 map, while the \textit{right column} shows those for the E24 map. From top to bottom, the rows show the crossing frequency, the characteristic crossing duration, and the residence time fraction. Different line styles correspond to different set-ups, as described in the legend.}  
    \label{fig:initialConditionTest}
\end{figure*}

For the map orientation test, we reorient the map randomly each time the Sun re-enters the map volume during the integration. This test is designed to assess how sensitive the encounter statistics are to the large scale organization of the ISM. The E24 map is particularly useful for this purpose, because it covers a sufficiently large volume of the Solar neighborhood to include linear structures \citep[see e.g.,][]{Kormann2026}. We find that the encounter statistics are not significantly affected by the map orientation, as shown in Fig.~\ref{fig:initialConditionTest}. This suggests that, for the statistical quantities considered here, the local ISM organization into clouds, the gas distribution, and the Solar orbit are the main drivers of the results, rather than the absolute orientation of large scale structures within the map. This conclusion is valid for the local Galactic volume probed by the present maps, but may not hold on larger scales, where the orientation of coherent and larger structures such as spiral arms might become important.

For the Solar initial condition tests, we estimated the encounter statistics using two different setups. In the first case, we kept the Solar position as in the reference case described in Sect.~\ref{sect:methods-orbitInt}, but assumed a Solar velocity relative to the LSR of ($U_\odot,\, V_\odot,\, W_\odot$) = (10.1, 4.0, 6.7)$\,\mathrm{km}\,\mathrm{s}^{-1}$ \citep{Hogg2005}. As shown in Fig.~\ref{fig:initialConditionTest} (under \texttt{Sun initial condition 1}), the encounter statistics remain mostly unchanged. The main exception is the crossing duration in the L20 map, shown in the left middle panel, for which the median value is larger than in the reference case. This is likely due to the smaller Solar velocity adopted from \citet{Hogg2005} compared with the value from \citet{Schoenrich2010} used in the reference integration, which increases the time required to cross individual density structures.

In the second case, we changed the Solar position, adopting a Galactocentric radius of $R_{\odot} = 8.33 \, \mathrm{kpc}$ \citep{Gillessen2009} and a vertical position of $z_{\odot}= 27 \, \mathrm{pc}$ \citep{Chen2001}, while keeping the Solar velocity relative to the LSR fixed to the \citet{Schoenrich2010} value. Also in this case, shown in Fig.~\ref{fig:initialConditionTest} (under \texttt{Sun initial condition 2}), the resulting crossing statistics are consistent with those obtained in the reference case.

These tests show that the encounter statistics are robust against the Solar initial condition variations considered here. This is expected because these changes do not substantially modify the Solar orbit: in all cases, the Sun samples a similar range of vertical heights, where the ISM vertical stratification does not vary strongly, and the Sun velocity changes only moderately. This is also in agreement with the study by \citet{Kokaia2019}, where they show that the $xy$ kinematic components of the stars do not affect the clouds hit rate.

\begin{figure*}
    \centering
    \includegraphics[width=1\textwidth]{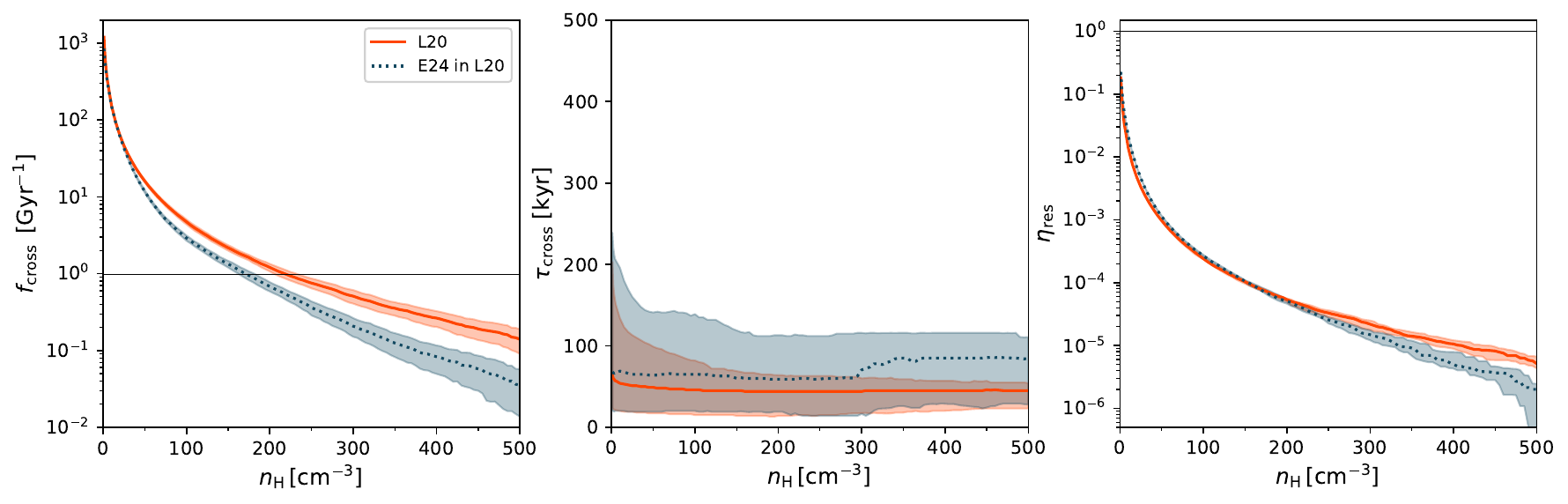}
    \caption{Same as Fig.~\ref{fig:crossing_statistics}, but for the case in which the E24 map is limited to the region covered by the L20 map.} 
    \label{fig:crossing_statistics_E24inL20}
\end{figure*}

We also tested the case in which the E24 volume is restricted to the volume covered by the L20 map, in order to verify whether the encounter statistics derived from the two maps become more consistent. As a first approximation, we repeated the analysis using only the E24 bins whose geometric centers lie within the L20 map boundaries. In this test, the E24 bins are still affected by the variation of bin volume with distance from the map center, although this effect is reduced. The radial size of the selected E24 bins ranges from $0.4$ to $3.3\,{\rm pc}$.
As shown in Fig.~\ref{fig:crossing_statistics_E24inL20}, the encounter statistics derived from the two maps become more similar when the same volume is considered, although differences remain. The agreement improves in particular for densities below $\sim100\,{\rm cm}^{-3}$, where the density regimes are equally well resolved in L20 and E24. This test therefore provides a useful consistency check, showing that the differences between the two maps are not produced by errors in the analysis pipeline, but mainly reflect their spatial discretizations.

\FloatBarrier

\section{Vertical stratification of the E24 map}\label{sect:E24_stratification}

In this section, we analyze the vertical stratification of the E24 map, following the same approach used for the L20 map in Sect.~\ref{sect:results-ISM_stratification}. Since the detailed vertical distribution of dust is not the main focus of this work, we perform a limited analysis.

\begin{figure*}
    \centering
    \includegraphics[width=0.95\textwidth]{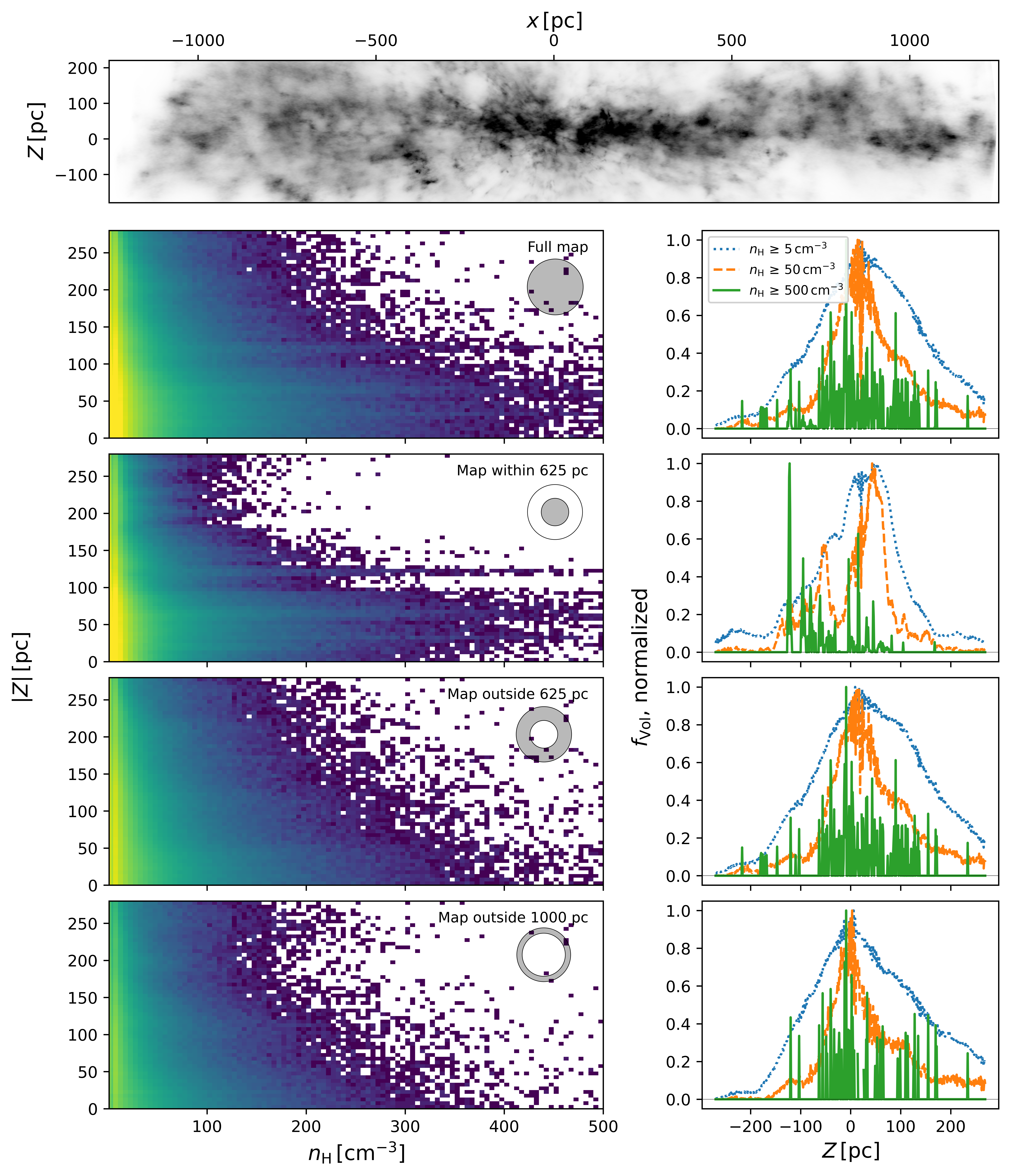}
    \caption{Vertical stratification of the ISM in the E24 map. 
    \textit{Top panel}: side projection of the E24 map.
    \textit{Left panels}: $z$-distribution of the density bins.
    \textit{Right panels}: normalized volume filling factor, $f_{\rm{Vol}}$, as a function of distance from the Galactic mid-plane and for different density thresholds. 
    Each row corresponds to a different region of the map (highlighted by the gray area in the circle), illustrating how the density resolution varies with distance from the map center.}  
    \label{fig:E24_zStratification}
\end{figure*}

Figure~\ref{fig:E24_zStratification} shows the side projection of the E24 map, together with the vertical distribution of the density bins and the volume filling factor for different density thresholds. We show these quantities for different radial selections of the map: the full E24 volume, the region within $625\,{\rm pc}$, the region beyond $625\,{\rm pc}$, and the region beyond $1000\,{\rm pc}$. These distance cuts are adopted because the E24 map covers a large volume of the Solar neighborhood and is therefore increasingly affected by distance dependent resolution effects.

We find that, in all four cases, the low density gas follows the expected decrease with increasing $|z|$. For higher density gas, the behavior is more complex. As found for the L20 map, the E24 map restricted to within $625\,{\rm pc}$ shows a distribution that cannot be described by a simple monotonic decline. Dense regions appear to be distributed relatively uniformly within $|z|\lesssim100-130\,{\rm pc}$, before declining rapidly at larger heights.
This behavior is not recovered in the outer regions of the E24 map, where the dense gas distribution becomes more strongly centered around $z=0$. In these outer regions, however, the highest density structures are also largely absent, as shown by the vertical distributions of the density bins. This is likely a consequence of the loss of spatial resolution with distance, which smooths compact dense structures and reduces their peak reconstructed densities.

A more detailed study of how the dust scale height depends on density threshold would therefore be valuable. However, such analysis is beyond the scope of the present work.

\FloatBarrier

\section{Height dependence of the incident flux and dust accretion}
\label{app:flux_massAccretion_zDependency}

\begin{figure*}
    \centering
    \includegraphics[width=0.95\textwidth]{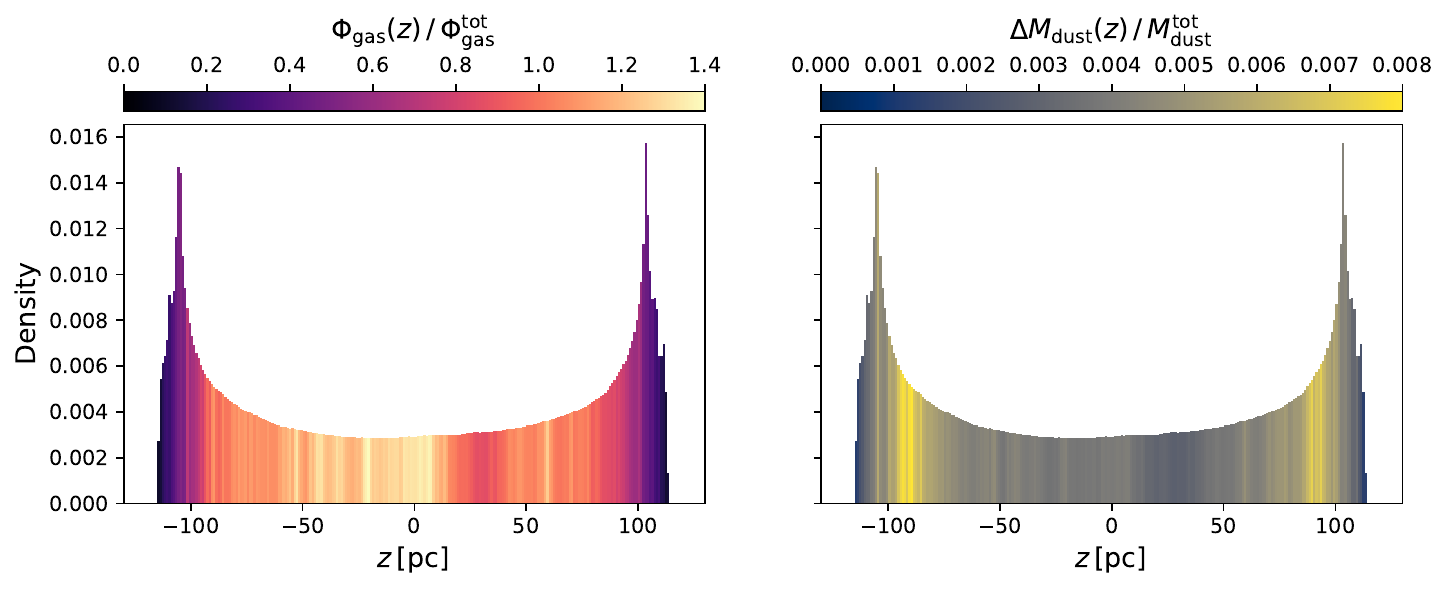}
    \caption{Histograms of the time spent by the Solar System at different Galactic vertical heights. The two panels show the same normalized distribution, but are color coded by different quantities. \textit{Left panel}: ratio between the mean instantaneous gas flux at each $z$ and the orbit averaged gas flux. \textit{Right panel}: fractional contribution of each height to the total accreted dust mass.} 
    \label{fig:flux_massAccreation_zDependency}
\end{figure*}

In Fig.~\ref{fig:flux_massAccreation_zDependency}, we show how the incident gas flux and the inferred accreted dust mass depend on the Galactic vertical height sampled by the Sun. The histogram shows the normalized distribution of time spent by the Sun at each height. The two panels show the same height distribution, but are color coded by different quantities.

In the left panel, the colors indicate the mean instantaneous gas flux in each vertical bin, normalized by the orbit averaged gas flux. This quantity measures the typical flux experienced by the Sun when it is located at a given height, independently of how much time the Sun spends there. It therefore reflects the combined effect of the gas density sampled at that height and the relative velocity between the Sun and the map.

In the right panel, the colors indicate the fractional contribution of each vertical bin to the total accreted dust mass. For fixed dust to gas ratio, effective cross section, and filtering efficiency, this quantity is equivalent to the fraction of the total integrated gas flux accumulated in each height bin. It therefore depends not only on the instantaneous gas flux, but also on the residence time of the Sun at that height.

This comparison illustrates why the regions of largest instantaneous exposure do not necessarily dominate the total accumulated dust mass. In particular, the instantaneous flux is enhanced closer to the Galactic midplane, whereas the total accreted mass receives a larger contribution from heights where the Sun spends more time, close to the vertical turning points of its orbit.

\end{appendix} 

\end{document}